\documentclass[submission,letterpaper,Phys]{SciPost}
\pdfoutput=1
\usepackage{amsmath,amssymb,mathtools,xspace}
\usepackage{amsthm}
\usepackage{booktabs,multirow,graphicx,tabularx,slashed}
\usepackage{hyperref}
\usepackage{color,xcolor}
\usepackage[normalem]{ulem}
\usepackage{enumitem}
\usepackage{feynmp}
\usepackage{braket}
\usepackage{tikz}
\usepackage{booktabs}
\usepackage{dsfont}
\usepackage{prettyref}
\usepackage{bbm}
\usepackage{comment}
\usepackage[bottom]{footmisc}

\usepackage{pifont}

\let\origref\ref
\newrefformat{sec}{\hyperref[#1]{Section~\origref*{#1}}}
\newrefformat{tab}{\hyperref[#1]{Table~\origref*{#1}}}
\newrefformat{fig}{\hyperref[#1]{Figure~\origref*{#1}}}
\newrefformat{eq}{eq.\,\hyperref[#1]{(\origref*{#1})}}
\newrefformat{al}{\hyperref[#1]{Algorithm~\origref*{#1}}}
\AtBeginDocument{\renewcommand*{\ref}[1]{\prettyref{#1}}}

\definecolor{EmeraldGreen}{HTML}{1ea78d}
\definecolor{EnglishRed}{HTML}{b02427}
\hypersetup{colorlinks=true,urlcolor=EmeraldGreen,citecolor=EmeraldGreen,linkcolor=EnglishRed}

\makeatletter
\@ifundefined{pdfoutput}{}{\DeclareGraphicsRule{*}{mps}{*}{}}
\makeatother

\definecolor{Gcolor}{HTML}{3b528b}
\definecolor{Dcolor}{HTML}{e41a1c}

\setitemize{itemsep=2pt,topsep=2pt,parsep=0pt,partopsep=0pt,labelindent=2pt,leftmargin=*}
\setenumerate{itemsep=0pt,topsep=2pt,parsep=0pt,partopsep=0pt,labelindent=3pt,leftmargin=*}

\usepackage{tikz}
\usetikzlibrary{shapes.geometric, arrows}
\usepackage{relsize}

\input{all.tikzdefs}

\tikzstyle{blank}=[fill=white, shape=circle, draw=white, inner sep=0.8pt]
\tikzstyle{dot}=[fill=black, shape=circle, draw=black, inner sep=0.8pt]
\tikzstyle{fat}=[fill=white, shape=circle, draw={rgb,255: red,176; green,36; blue,39}, dashed, line width=1pt, inner sep=0.8pt]

\tikzstyle{edge}=[-, draw=PropagatorColor, line width=1.2pt, line cap=rect, preaction={{draw=white,line width=2.5pt}}]
\tikzstyle{incoming edge}=[line width=1pt, line cap=rect, draw={rgb,255: red,102; green,102; blue,102}, {|-}]
\tikzstyle{outgoing edge}=[line width=1pt, line cap=rect, draw={rgb,255: red,102; green,102; blue,102}, ->]
\tikzstyle{external edge}=[line width=1pt, line cap=rect, draw={rgb,255: red,102; green,102; blue,102}, -]
\tikzstyle{top}=[-, draw=TopPropagatorColor, line width=1.9pt, line cap=rect, preaction={{draw=white,line width=2.5pt}}]
\tikzstyle{edge dot1}=[-, postaction=decorate, decoration={markings,mark=at position .50 with {\node[style=dot]{};}}]
\tikzstyle{edge dot2}=[-, postaction=decorate, decoration={markings,mark=between positions 0.33 and 0.67 step 0.33 with {\node[style=dot]{};}}]
\tikzstyle{edge dot3}=[-, postaction=decorate, decoration={markings,mark=between positions 0.25 and 0.76 step 0.25 with {\node[style=dot]{};}}]
\tikzstyle{edge dot4}=[-, postaction=decorate, decoration={markings,mark=between positions 0.20 and 0.81 step 0.20 with {\node[style=dot]{};}}]
\tikzstyle{dot1}=[-, draw=none, postaction=decorate, decoration={markings,mark=at position .50 with {\node[style=dot]{};}}]
\tikzstyle{dot2}=[-, draw=none, postaction=decorate, decoration={markings,mark=between positions 0.33 and 0.67 step 0.33 with {\node[style=dot]{};}}]
\tikzstyle{dot3}=[-, draw=none, postaction=decorate, decoration={markings,mark=between positions 0.25 and 0.76 step 0.25 with {\node[style=dot]{};}}]
\tikzstyle{dot4}=[-, draw=none, postaction=decorate, decoration={markings,mark=between positions 0.20 and 0.81 step 0.20 with {\node[style=dot]{};}}]
\tikzstyle{incoming}=[line width=1pt, line cap=rect, draw=LegColor, {|-}]
\tikzstyle{outgoing}=[line width=1pt, line cap=rect, draw=LegColor, ->]
\tikzstyle{outgoing top}=[line width=1pt, line cap=rect, draw=TopLegColor, ->]
\tikzstyle{outgoing higgs}=[line width=1pt, line cap=rect, draw=HiggsLegColor, ->]
\tikzstyle{edge}=[-, draw={rgb,255: red,176; green,36; blue,39}, line width=1pt, preaction={{draw=white,line width=2pt}}, line cap=rect]
\tikzstyle{massive edge}=[-, draw={rgb,255: red,11; green,60; blue,191}, line width=2.0pt, preaction={{draw=white,line width=2.5pt}}, line cap=rect]
\tikzstyle{massive2 edge}=[-, draw={rgb,255: red,59; green,196; blue,114}, line width=2.0pt, preaction={{draw=white,line width=2.5pt}}, line cap=rect]
\tikzstyle{cut edge}=[-, draw={rgb,255: red,64; green,64; blue,64}, line width=0.5pt, densely dashed, line cap=rect]
\tikzstyle{xcut edge}=[-, draw={rgb,255: red,64; green,64; blue,64}, line width=0.5pt, densely dashed, line cap=rect, postaction={decorate, decoration={markings,mark=at position .50 with {\node[cross out,solid,draw=white,line width=2pt,inner sep=1.4pt,transform shape] {};}}}, postaction={decorate, decoration={markings,mark=at position .50 with {\node[cross out,solid,draw={rgb,255: red,176; green,36; blue,39},line width=1pt,inner sep=1.8pt,transform shape] {};}}}]
\tikzstyle{xcut edge 1/3}=[-, draw={rgb,255: red,64; green,64; blue,64}, line width=0.5pt, densely dashed, line cap=rect, postaction={decorate, decoration={markings,mark=at position .33 with {\node[cross out,solid,draw=white,line width=2pt,inner sep=1.4pt,transform shape] {};}}}, postaction={decorate, decoration={markings,mark=at position .33 with {\node[cross out,solid,draw={rgb,255: red,176; green,36; blue,39},line width=1pt,inner sep=1.8pt,transform shape] {};}}}]
\tikzstyle{cut}=[-, draw={rgb,255: red,61; green,171; blue,83}, line width=0.7pt, dotted, line cap=rect]
\tikzstyle{photon}=[-, draw={rgb,255: red,176; green,36; blue,39}, line width=1pt, preaction={{draw=white,line width=2pt}}, line cap=rect, decorate, decoration=snake]
\tikzstyle{gluon}=[-, draw={rgb,255: red,176; green,36; blue,39}, line width=1pt, preaction={{draw=white,line width=2pt}}, line cap=rect, decorate, decoration={coil,aspect=1.4,segment length=2.5mm}]
\tikzstyle{gluoncoil}=[-, decorate, decoration={coil,aspect=1.4,segment length=2.5mm}]
\tikzstyle{fermion}=[-, draw={rgb,255: red,176; green,36; blue,39}, line width=1pt, preaction={{draw=white,line width=2pt}}, line cap=rect, postaction=decorate, decoration={markings,mark=at position .60 with {\arrow{stealth[round]}}}]
\tikzstyle{ghost}=[-, style=fermion, line width=1pt, line cap=round, dash pattern={on 0pt off 3\pgflinewidth}]
\tikzstyle{scalar}=[-, line width=1pt, densely dashed, draw={rgb,255: red,102; green,102; blue,102}]
\tikzstyle{fermionarrow}=[-,postaction=decorate, decoration={markings,mark=at position .60 with {\arrow{stealth[round]}}}]
\tikzstyle{arrow}=[{-{Classical TikZ Rightarrow[length=2mm,width=1.5mm]}}, draw={rgb,255: red,61; green,171; blue,83}, line width=1pt, preaction={{draw=white,line width=2pt}}, line cap=rect]
\tikzstyle{brace}=[-,draw={rgb,255: red,61; green,171; blue,83}, line width=1pt, decorate, decoration={brace,amplitude=5pt}]

\tikzstyle{generator} = [rectangle, rounded corners, minimum width=3cm, minimum height=1cm,text centered, draw=Gcolor]
\tikzstyle{discriminator} = [rectangle, rounded corners, minimum width=3cm, minimum height=1cm,text centered, draw=Dcolor]
\tikzstyle{io} = [circle, trapezium left angle=70, trapezium right angle=110, minimum width=1cm, minimum height=1cm, text centered, draw=black]

\tikzstyle{process} = [rectangle, minimum width=1cm, minimum height=1cm, text centered, draw=black]
\tikzstyle{decision} = [rectangle, minimum width=1cm, minimum height=1cm, text centered, draw=black]

\tikzstyle{arrow} = [thick,->,>=stealth]

\usepackage{algorithm}
\usepackage{algorithmic}

\usepackage{xcolor}

\def\rd{\mathrm{d}}
\def\imag{\mathrm{i}}
\def\pysecdec{py\textsc{SecDec}}

\newcommand{\noun}[1]{{\textsc{#1}}}

\newcommand\one{\leavevmode\hbox{\small1\normalsize\kern-.33em1}}

\def\slashchar#1{\setbox0=\hbox{$#1$}           
   \dimen0=\wd0                                 
   \setbox1=\hbox{/} \dimen1=\wd1               
   \ifdim\dimen0>\dimen1                        
      \rlap{\hbox to \dimen0{\hfil/\hfil}}      
      #1                                        
   \else                                        
      \rlap{\hbox to \dimen1{\hfil$#1$\hfil}}   
      /                                         
   \fi}

\newcommand{\ie}{\textsl{i.e.}\;}

\renewcommand{\d}{{\text{d}}}

\newcommand{\eps}{\epsilon}

\DeclareMathOperator*{\diag}{diag}

\begin{document}


\begin{center}{\Large \textbf{
Targeting Multi-Loop Integrals with Neural Networks
}}\end{center}

\begin{center}
Ramon Winterhalder\textsuperscript{1,2,3},
Vitaly Magerya\textsuperscript{4},
Emilio Villa\textsuperscript{4}, 
Stephen P. Jones\textsuperscript{5},\\
Matthias Kerner\textsuperscript{4,6},
Anja Butter\textsuperscript{1,2},
Gudrun Heinrich\textsuperscript{2,4},
Tilman Plehn\textsuperscript{1,2}

\end{center}
\begin{center}
{\bf 1} Institut f\"ur Theoretische Physik, Universit\"at Heidelberg, Germany\\
{\bf 2} HEiKA – Heidelberg Karlsruhe Strategic Partnership, Heidelberg University, Karlsruhe Institute of Technology (KIT), Germany\\
{\bf 3} Centre for Cosmology, Particle Physics and Phenomenology (CP3), \\Universit\'e catholique de Louvain, Belgium\\
{\bf 4} Institut f\"ur Theoretische Physik, Karlsruher Institut f\"ur Technologie, Germany\\
{\bf 5} Institute for Particle Physics Phenomenology, Durham University, UK\\
{\bf 6} Institut f\"ur Astroteilchenphysik, Karlsruher Institut f\"ur Technologie, Germany
\end{center}
\begin{center}
\today
\end{center}


\section*{Abstract}
         {\bf Numerical evaluations of Feynman integrals
         often proceed via a deformation of the integration contour into the complex plane. While valid contours are easy to construct, the numerical precision for a multi-loop integral can depend critically on the chosen contour. We present methods to optimize this contour using a combination of optimized, global complex shifts and a normalizing flow. They can lead to a significant gain in precision.}

\vspace{10pt}
\noindent\rule{\textwidth}{1pt}
\tableofcontents\thispagestyle{fancy}
\noindent\rule{\textwidth}{1pt}

\newpage
\section{Introduction}
\label{sec:intro}

High-precision predictions based on quantum field theory are the cornerstone in the LHC research program probing fundamental physics and the way towards identifying physics beyond the Standard Model. Improved analysis techniques, better controlled systematics and the planned 25-fold luminosity increase during the LHC Runs~3 and~4 translate into a major challenge for the corresponding theory calculations and
simulations~\cite{Heinrich:2020ybq}. Furthermore, the prospect of a future lepton collider will require the calculation of electroweak corrections with many different mass scales, where the underlying multi-loop integrals are out of reach for analytical approaches so far, while numerical methods are promising~\cite{Freitas:2021oiq}.

For all aspects of LHC simulations, including numerical approaches to loop integrals and amplitudes, computing
time and algorithmic efficiency are essential.  
Therefore, it is crucial that we investigate new methods which have the potential to improve the numerical efficiency, in particular machine learning (ML) methods.
Along the 
established LHC simulation chain, machine learning has already shown great promise when it comes to faster and more precise
predictions~\cite{Butter:2020tvl}. This includes phase space
integration~\cite{maxim,Chen:2020nfb}, phase space
sampling~\cite{Bothmann:2020ywa,Gao:2020vdv,Gao:2020zvv,Danziger:2021eeg},
amplitude evaluation~\cite{Bishara:2019iwh,Badger:2020uow,Aylett-Bullock:2021hmo,Maitre:2021uaa}, event
subtraction~\cite{Butter:2019eyo}, event
unweighting~\cite{Verheyen:2020bjw,Backes:2020vka}, parton
showering~\cite{shower,locationGAN,juniprshower,Dohi:2020eda}, parton
densities~\cite{NNPDF:2014otw,NNPDF:2019ubu} or particle flow
descriptions~\cite{DiBello:2020bas,Baldi:2020hjm}. Full neural network-based
event
generators~\cite{dutch,gan_datasets,DijetGAN2,Butter:2019cae,Alanazi:2020klf,Butter:2021csz}
can be used to invert the simulation chain and unfold detector effects
as well as QCD jet
radiation~\cite{Datta:2018mwd,fcgan,Bellagente:2020piv}.  
An important issue in applying machine learning to LHC
predictions is uncertainty control and
quantification, which is being addressed with increasing success~\cite{Badger:2020uow,Diefenbacher:2020rna,Bellagente:2021yyh,Winterhalder:2021ave,Danziger:2021eeg,Butter:2021csz}.

Essentially all of these ML-applications are driven by three
properties of neural networks: they are very flexible in what they
describe and how they are trained, they provide an excellent
interpolation, and they are extremely fast once trained. 
These numerical advantages naturally lead us to investigate where ML could be useful in multi-loop calculations. 

We present a first application of modern
neural networks in loop integrals.
Our starting points are Feynman integrals in a parametric representation, where the loop momenta have been integrated out analytically by the standard procedure, leading to the two Symanzik polynomials ${\cal U}$ and ${\cal F}$, see e.g.~\cite{Heinrich:2020ybq} for a description.
Such integrals often have poles which manifest themselves as powers of $1/\eps$ in dimensional regularization and can be factorized efficiently with sector decomposition~\cite{Binoth:2000ps,Heinrich:2008si}.
After factorizing the poles, integrable singularities, related for example to thresholds, remain. They can be dealt with by a deformation of the integration contour into the complex plane~\cite{Soper:1999xk,Binoth:2005ff,Nagy:2006xy,Anastasiou:2007qb,Lazopoulos:2007ix,Beerli:2008zz}.
An automated procedure to do so has been implemented for the first time in {\sc SecDec}~\cite{Borowka:2012yc} and has been refined in {\sc SecDec}-3~\cite{Borowka:2015mxa} and \pysecdec~\cite{Borowka:2017idc,Borowka:2018goh,Heinrich:2021dbf}.
 
The deformation of the integration contour can be performed in many ways, the only requirement is that no pole is crossed by the deformation. Applications to multi-loop integrals with a certain complexity show that the numerical precision can vary by orders of magnitude depending on the choice of a particular contour. In this work we present methods to optimize the choice of the contour based on neural networks.

In \ref{sec:feynman_integrals} we briefly review the construction of the Feynman parametric representation of multi-loop integrals, which forms the starting point of our investigations, as well as the contour deformation procedure employed in \pysecdec. In \ref{sec:ml_approach} we describe our new approach to contour deformation based on neural networks and show results for several examples, before we give an Outlook.

\section{Multi-loop Feynman integrals}
\label{sec:feynman_integrals}

For our study of ML-methods we focus on the numerical evaluation of
integrals in the Feynman-parameter representation. Before we show how
neural networks can improve the numerical evaluation of such
integrals, we briefly review their definition and the way they are
evaluated in \pysecdec{}.

\subsubsection*{Feynman parametrization}

A generic scalar Feynman integral in $D$ space-time
dimensions with $L$ loops and $N$ propagators
of arbitrary powers $\nu_j$ can be represented by
\begin{align}\label{eq:Feynman}
G &=
\int\limits_{-\infty}^{\infty} \left(\prod\limits_{l=1}^{L} \frac{\d^D k_l}{i\pi^{\frac{D}{2}}}\right)
\prod\limits_{j=1}^{N} 
\frac{1}{P_{j}^{\nu_j}(\{k\},\{p\},m_j^2)}\;, 
\end{align}
where the propagators $P_{j}$ are of the form
$P_{j}(\{k\},\{p\},m_j^2)=q_j^2-m_j^2+\imag\delta$, with $q_j$ being a
linear combination of loop momenta $k$ and external momenta $p$.
Introducing Feynman parameters $x_j$ through
\begin{align}\label{eq:Feynmanpara}
  \prod\limits_{j=1}^{N} \frac{1}{P_{j}^{\nu_j}}&
  =\frac{\Gamma(\nu)}{\prod\limits_{j=1}^N\Gamma(\nu_j)}\int\limits_0^\infty \left(\prod\limits_{j=1}^N\,\d x_j\,x_j^{\nu_j-1}\right)
  \frac{\delta\!\left(1-\sum_{i=1}^N x_i \right)}{\left( \sum_{j=1}^Nx_jP_j \right)^{\nu}},
  \quad \text{with} \quad \nu\equiv\sum_{j=1}^N\nu_j,
\end{align}
leads to
\begin{align}
\begin{split}
G &= \frac{\Gamma(\nu)}{\prod\limits_{j=1}^{N}\Gamma(\nu_j)}\int\limits_0^\infty \left(\prod\limits_{j=1}^{N}\,\d x_j\,x_j^{\nu_j-1}\right)
 \delta\!\left( 1-\sum_{i=1}^N x_i \right) \times\\
&\times \int\limits_{-\infty}^{\infty} \left(\prod\limits_{l=1}^{L} \frac{\d^Dk_l}{i\pi^{\frac{D}{2}}}\right) 
\left[ 
       \sum\limits_{j,l=1}^{L} k_j\cdot k_l \, M_{jl} - 
      2\sum\limits_{j=1}^{L} k_j\cdot Q_j +J +i\delta\right]^{-\nu}.
\end{split}
\end{align}
Further details can be found e.g. in \cite{Heinrich:2020ybq,Binoth:2000ps}.
Integration over the momenta gives us an expression in terms of the Symanzik
polynomials $\mathcal{U}$ and $\mathcal{F}$,
\begin{align}
\begin{split}
  G = \frac{(-1)^{\nu}\, \Gamma(\nu-LD/2)}{\prod_{j=1}^{N}\Gamma(\nu_j)}
 \int\limits_{0}^{\infty} \left(\prod\limits_{j=1}^{N}\d x_j\,x_j^{\nu_j-1}\right)\,\delta\!\left(1-\sum_{l=1}^N x_l\right)
 \frac{\mathcal{U}^{\nu-(L+1) D/2}}{{\mathcal F}^{\nu-L D/2}} \; , \\
 \text{with} \qquad 
{\cal U} \equiv \det(M)
 \qquad \text{and} \qquad 
{\cal F} \equiv \det(M) \left[ \sum_{i,j=1}^L Q_i \left(M^{-1}\right)_{ij}Q_j-J-\imag\delta \right].
\label{eq:feynint}
\end{split}
\end{align}
The first Symanzik polynomial, ${\cal U}$, is a positive semi-definite
function of the Feynman parameters. The second Symanzik polynomial,
${\cal F}$, contains kinematic invariants and Feynman parameters. A
vanishing ${\cal F}$ is a necessary, but not sufficient condition
for infrared or kinematic singularities to arise.


\subsubsection*{Contour deformation in \pysecdec{}}

After mapping all integration variables onto the unit-hypercube and
integrating out the delta distribution we can absorb any additional
factors by redefining and renaming $\mathcal{U}\to U$ and
$\mathcal{F}\to F$. Then, \ref{eq:feynint} can be written in the
compact form
\begin{align}
G=\int\limits_{0}^{1} \prod\limits_{j=1}^{N-1}\d x_j\,x_j^{\nu_j-1}\,
\frac{U^{\nu-(L+1) D/2}}{F^{\nu-L D/2}}
=\int\limits_{0}^{1} \prod\limits_{j=1}^{N-1}\d x_j \; \mathcal{I}(\vec{x})\; ,
\label{eq:feynman_integral}
\end{align}
where, again, $F(\vec{x})$ can vanish inside the integration
region. If we deform the integration over a Feynman parameter away
from the real line segment $x \in [0, 1]$ into the complex $z$-plane,
Cauchy’s theorem ensures that the integral does not change as long as
no singularities are enclosed by the contour,
\begin{align}
\begin{split}
 0&=\oint_c  \prod\limits_{j=1}^{N}\rd z_j  \, {\cal I}(\vec{z}) 
 =  \int\limits_{0}^{1} \prod\limits_{j=1}^{N}\rd x_j \, {\cal I}(\vec{x})
   +\int_\gamma \prod\limits_{j=1}^{N}\rd z_j \, {\cal I}(\vec{z}) \\
    \Leftrightarrow \quad &
    \int\limits_{0}^{1}  \prod\limits_{j=1}^{N}\rd x_j\, \mathcal{I}(\vec{x})
    =-\int_\gamma \prod\limits_{j=1}^{N}\rd z_j\, {\cal I}(\vec{z}) 
    =\int\limits_{0}^{1}  \prod\limits_{j=1}^{N}\rd x_j\;
    \det\!\left(\frac{\partial \vec{z}(\vec{x})}{\partial \vec{x}}\right)
               {\cal I}(\vec{z}(\vec{x}))\;.   
\end{split}
\label{eq:contour}
\end{align}
In the complex plane the $-\imag\delta$ prescription in
\ref{eq:Feynman} and \ref{eq:feynint} ensures that we stay on the
physical and causal Riemann sheet.

To construct an appropriate deformation into the complex plane we
write $\vec{z}=\vec{x}-\imag\vec{\tau}$ and expand $F(\vec{z})$ around
$\vec{x}$,
\begin{align}
\label{eq:f-expansion}
F(\vec{z})=F(\vec{x})
    -\imag\sum_j \tau_j \frac{\partial F(\vec{x})}{\partial x_j}
    -\frac{1}{2} \sum_{j,k} \tau_j \tau_k \frac{\partial^2 F(\vec{x})}{\partial x_j \partial x_k}
    +\frac{\imag}{6} \sum_{j,k,l} \tau_j \tau_k \tau_l \frac{\partial^3 F(\vec{x})}{\partial x_j \partial x_k \partial x_l}
    +\mathcal{O}(\tau^4) \; .
\end{align}
The second term gives the leading term for the imaginary part of
$F(\vec{x})$. We can guarantee that it is always negative by choosing
$\tau_j\propto\partial F(\vec{x})/\partial x_j$. We also ensure that
the integration endpoints are invariant under the contour deformation
by requiring $\tau_j \propto x_j (1-x_j)$, defining the contour
deformation as
\begin{align}
\label{eq:deformp}
\tau_j = \lambda_j x_j(1-x_j)\frac{\partial F(\vec{x})}{\partial x_j},
\qquad \text{with} \qquad \lambda_j > 0\;.
\end{align}
The \textit{deformation parameters} $\lambda_j$ can be chosen
arbitrarily, provided they are small enough for the leading order in
$\tau$ to dominate the imaginary part of $F(\vec{z})$. If the $\lambda_j$ are too
large the tri-linear term in \ref{eq:f-expansion} can flip the sign of the imaginary part.

In \noun{SecDec~3.0}~\cite{Borowka:2015mxa}, the $\lambda_j$ are chosen by first determining their maximal values at which the tri-linear terms in \ref{eq:f-expansion} have the same magnitude as the linear ones; then, some fractions of these maximal values are selected via several heuristics depending on the sampled values of $\partial F(\vec{x})/\partial x_j$; a detailed description is provided in Section~6.2.3 of Ref.~\cite{Borowka:2014aaa}. In \pysecdec{}~\cite{Borowka:2017idc} $\lambda_j$ are selected as the smallest of sampled values of 
\begin{align}
    \left|\, x_j (1-x_j) \frac{\partial F(\vec{x})}{\partial x_j} \,\right|^{-1} .
\end{align}
In both cases the initial selection is followed by iterative refinement steps: if during the integration a sign check error occurs (\ie either $\mathrm{Im}\,F(\vec{x})$ is found to be positive or $\mathrm{Re}\, U(\vec{x})$ is found to be negative) for one of the sampling points, then all $\lambda_j$ are multiplied by a factor of $0.9$ and the integration is repeated.  As a consequence, \pysecdec{}
often selects the largest allowed $\lambda$ vector along the initially
chosen direction.

\subsubsection*{Example diagrams}

\begin{figure}[t]
  \centering
  \raisebox{0.5ex}{\begin{tikzpicture}
	\begin{pgfonlayer}{nodelayer}
		\node [style=dot] (0) at (1, 0.75) {};
		\node [style=dot] (1) at (-0.25, 0.75) {};
		\node [style=dot] (2) at (-0.25, -0.75) {};
		\node [style=dot] (3) at (1, -0.75) {};
		\node [style=dot] (4) at (2, 0) {};
		\node [style=none] (5) at (1.75, 0.75) {};
		\node [style=none] (6) at (-1, 0.75) {};
		\node [style=none] (7) at (-1, -0.75) {};
		\node [style=none] (8) at (1.75, -0.75) {};
		\node [style=none] (9) at (2.75, 0) {};
		\node [style=none] (10) at (-1.25, -0.75) {$p_2$};
		\node [style=none] (11) at (-1.25, 0.75) {$p_1$};
		\node [style=none] (12) at (2, 0.75) {$p_3$};
		\node [style=none] (13) at (3, 0) {$p_4$};
		\node [style=none] (14) at (2, -0.75) {$p_5$};
		\node [style=none] (20) at (1.25, -0.25) {$m$};
		\node [style=none] (21) at (1.25, 1) {$m$};
		\node [style=none] (22) at (0.5, 0.5) {$m$};
		\node [style=none] (23) at (0.5, -0.5) {$m$};
		\node [style=none] (24) at (-0.5, 0) {$m$};
		\node [style=none] (25) at (2.25, -0.25) {$m$};
	\end{pgfonlayer}
	\begin{pgfonlayer}{edgelayer}
		\draw [style=massive2 edge] (8.center) to (3);
		\draw [style=edge] (0) to (4);
		\draw [style=massive edge] (5.center) to (0);
		\draw [style=massive edge] (9.center) to (4);
		\draw [style=massive edge] (0) to (1);
		\draw [style=massive edge] (1) to (2);
		\draw [style=massive edge] (2) to (3);
		\draw [style=massive edge] (3) to (4);
		\draw [style=edge] (6.center) to (1);
		\draw [style=edge] (7.center) to (2);
	\end{pgfonlayer}
\end{tikzpicture}}
  \hspace*{10mm}
  \raisebox{0.5ex}{\begin{tikzpicture}
	\begin{pgfonlayer}{nodelayer}
		\node [style=none] (0) at (-1.75, -0.75) {};
		\node [style=dot] (1) at (-1, -0.75) {};
		\node [style=dot] (2) at (-1, 0.75) {};
		\node [style=none] (3) at (-1.75, 0.75) {};
		\node [style=dot] (4) at (0, -0.75) {};
		\node [style=dot] (5) at (0, 0.75) {};
		\node [style=dot] (6) at (1, -0.75) {};
		\node [style=dot] (7) at (1, 0.75) {};
		\node [style=none] (8) at (1.75, -0.75) {};
		\node [style=none] (9) at (1.75, 0.75) {};
		\node [style=none] (10) at (-2, 0.75) {$p_1$};
		\node [style=none] (11) at (-2, -0.75) {$p_2$};
		\node [style=none] (12) at (2, -0.75) {$p_3$};
		\node [style=none] (13) at (2, 0.75) {$p_4$};
		\node [style=none] (14) at (-0.5, -0.5) {$m$};
		\node [style=none] (15) at (0.5, -0.5) {$m$};
		\node [style=none] (16) at (0.5, 0.5) {$m$};
		\node [style=none] (17) at (-0.5, 0.5) {$m$};
		\node [style=none] (18) at (-0.75, 0) {$m$};
		\node [style=none] (20) at (1.25, -0.5) {$m$};
	\end{pgfonlayer}
	\begin{pgfonlayer}{edgelayer}
		\draw [style=edge] (3.center) to (2);
		\draw [style=edge] (0.center) to (1);
		\draw [style=massive edge] (8.center) to (6);
		\draw [style=massive edge] (6) to (4);
		\draw [style=massive edge] (4) to (1);
		\draw [style=massive edge] (1) to (2);
		\draw [style=massive edge] (2) to (5);
		\draw [style=massive edge] (5) to (7);
		\draw [style=edge] (4) to (5);
		\draw [style=edge] (6) to (7);
		\draw [style=massive2 edge] (7) to (9.center);
	\end{pgfonlayer}
\end{tikzpicture}} \\[10mm]
  \raisebox{0.5ex}{\begin{tikzpicture}
	\begin{pgfonlayer}{nodelayer}
		\node [style=none] (0) at (2.75, 0) {};
		\node [style=dot] (1) at (2, 0) {};
		\node [style=dot] (2) at (1, 0.75) {};
		\node [style=dot] (3) at (1, -0.75) {};
		\node [style=dot] (4) at (-0.25, 0.75) {};
		\node [style=dot] (5) at (-0.25, -0.75) {};
		\node [style=none] (6) at (-1, 0.75) {};
		\node [style=none] (7) at (-1, -0.75) {};
		\node [style=none] (8) at (3, 0) {$p_3$};
		\node [style=none] (9) at (-1.25, 0.75) {$p_1$};
		\node [style=none] (10) at (-1.25, -0.75) {$p_2$};
		\node [style=none] (11) at (1.75, 0.5) {$m$};
		\node [style=none] (12) at (0.75, 0) {$m$};
		\node [style=none] (13) at (1.75, -0.5) {$m$};
	\end{pgfonlayer}
	\begin{pgfonlayer}{edgelayer}
		\draw [style=massive2 edge] (0.center) to (1);
		\draw [style=massive edge] (1) to (2);
		\draw [style=massive edge] (2) to (3);
		\draw [style=massive edge] (3) to (1);
		\draw [style=edge] (2) to (4);
		\draw [style=edge] (3) to (5);
		\draw [style=edge] (5) to (4);
		\draw [style=edge] (4) to (6.center);
		\draw [style=edge] (5) to (7.center);
	\end{pgfonlayer}
\end{tikzpicture}}
  \hspace*{10mm}
  \raisebox{0.5ex}{\begin{tikzpicture}
	\begin{pgfonlayer}{nodelayer}
		\node [style=dot] (0) at (-1, -0.75) {};
		\node [style=dot] (1) at (-1, 0.75) {};
		\node [style=dot] (2) at (0, -0.75) {};
		\node [style=dot] (4) at (1, -0.75) {};
		\node [style=dot] (5) at (1, 0.75) {};
		\node [style=none] (6) at (-1.75, -0.75) {};
		\node [style=none] (7) at (-1.75, 0.75) {};
		\node [style=none] (8) at (1.75, -0.75) {};
		\node [style=none] (9) at (1.75, 0.75) {};
		\node [style=none] (10) at (2, -0.75) {$p_4$};
		\node [style=none] (11) at (1.25, 0) {$m$};
		\node [style=none] (12) at (-1.25, 0) {$m$};
		\node [style=none] (13) at (0, 0.5) {$m$};
		\node [style=none] (14) at (-0.5, -0.5) {$m$};
		\node [style=none] (16) at (0.5, -0.5) {$m$};
		\node [style=none] (17) at (-2, 0.75) {$p_1$};
		\node [style=none] (18) at (-2, -0.75) {$p_2$};
		\node [style=none] (19) at (2, 0.75) {$p_3$};
	\end{pgfonlayer}
	\begin{pgfonlayer}{edgelayer}
		\draw [style=massive edge] (0) to (2);
		\draw [style=massive edge] (1) to (0);
		\draw [style=edge] (5) to (9.center);
		\draw [style=massive2 edge] (8.center) to (4);
		\draw [style=edge] (0) to (6.center);
		\draw [style=edge] (1) to (7.center);
		\draw [style=edge] (5) to (2);
		\draw [style=massive edge] (5) to (4);
		\draw [style=massive edge] (5) to (1);
		\draw [style=massive edge] (4) to (2);
	\end{pgfonlayer}
\end{tikzpicture}}
  \caption{Feynman diagrams for our four example integrals, which we call
    pentagon1L, ladder2L (first line) and triangle2L, 
    elliptic2L (second line). The blue lines denote massive lines, green lines denote massive or off-shell external legs (with a mass different from $m$).}
  \label{fig:feyn_diags}
\end{figure}
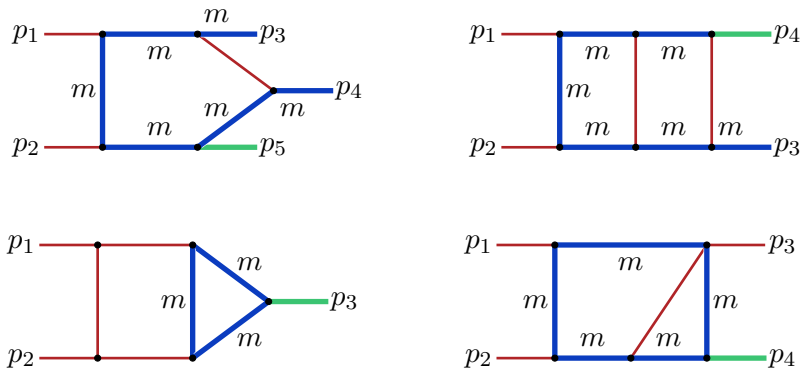

The Feynman diagrams we use to develop and benchmark our approaches are shown in \ref{fig:feyn_diags}.

The top left diagram is a one-loop pentagon integral as it occurs in the production of a top quark pair in association with another massive particle and depends on four independent Mandelstam invariants as well as the top quark mass and the invariant mass of $p_5$. Analytically it depends on logarithms and dilogarithms of ratios of kinematic invariants, leading to a complicated branch-cut structure. After Feynman parametrization the corresponding integral is described by 4 independent Feynman parameters.

The top right diagram is a two-loop box diagram with one massive on-shell leg and one off-shell leg. This diagram is a topology occurring for example in $t\bar{t}V$ production at two loops, where the boson $V$ is radiated off an external top quark. It is close to the configuration of a 2-loop gluon ladder diagram where the exchange of gluons between two top quark lines gives rise to a Coulomb singularity. 
The analytic expression for this type of diagram is not known, but it is anticipated that it will contain elliptic functions. This integral depends on 6 Feynman parameters and is the most complicated example we consider in terms of dimensionality.

The diagram on the lower left of \ref{fig:feyn_diags} is a two-loop three-point function with a massive sub-triangle occurring, for instance, in NLO corrections to Higgs production in gluon fusion.
It is the easiest 2-loop diagram we consider and serves as a stepping stone towards more complicated 2-loop diagrams. Analytic results for this diagram can be found in Refs.~\cite{Fleischer:1997bw,Davydychev:2003mv,Bonciani:2003hc}. Depending on 5 Feynman parameters this integral is in between the previous two examples in terms of dimensionality of the integration.

The diagram on the lower right is a topology occurring in Higgs+jet production in gluon fusion at two loops. Its analytic expression contains elliptic functions and therefore is cutting edge for integrals that are currently accessible analytically. It has been calculated (semi-)analytically in Refs.~\cite{Bonciani:2016qxi,Frellesvig:2019byn} and also served as a benchmark for the development of the program \pysecdec{}~\cite{Borowka:2017idc}, where it is contained in the list of examples. This integral is 5-dimensional, so it has the same number of Feynman parameters as the triangle diagram, but it depends on four kinematic invariants rather than two.

\section{Machine learning contour deformations}
\label{sec:ml_approach}

\begin{figure}[b!]
  \includegraphics[width=0.495\textwidth]{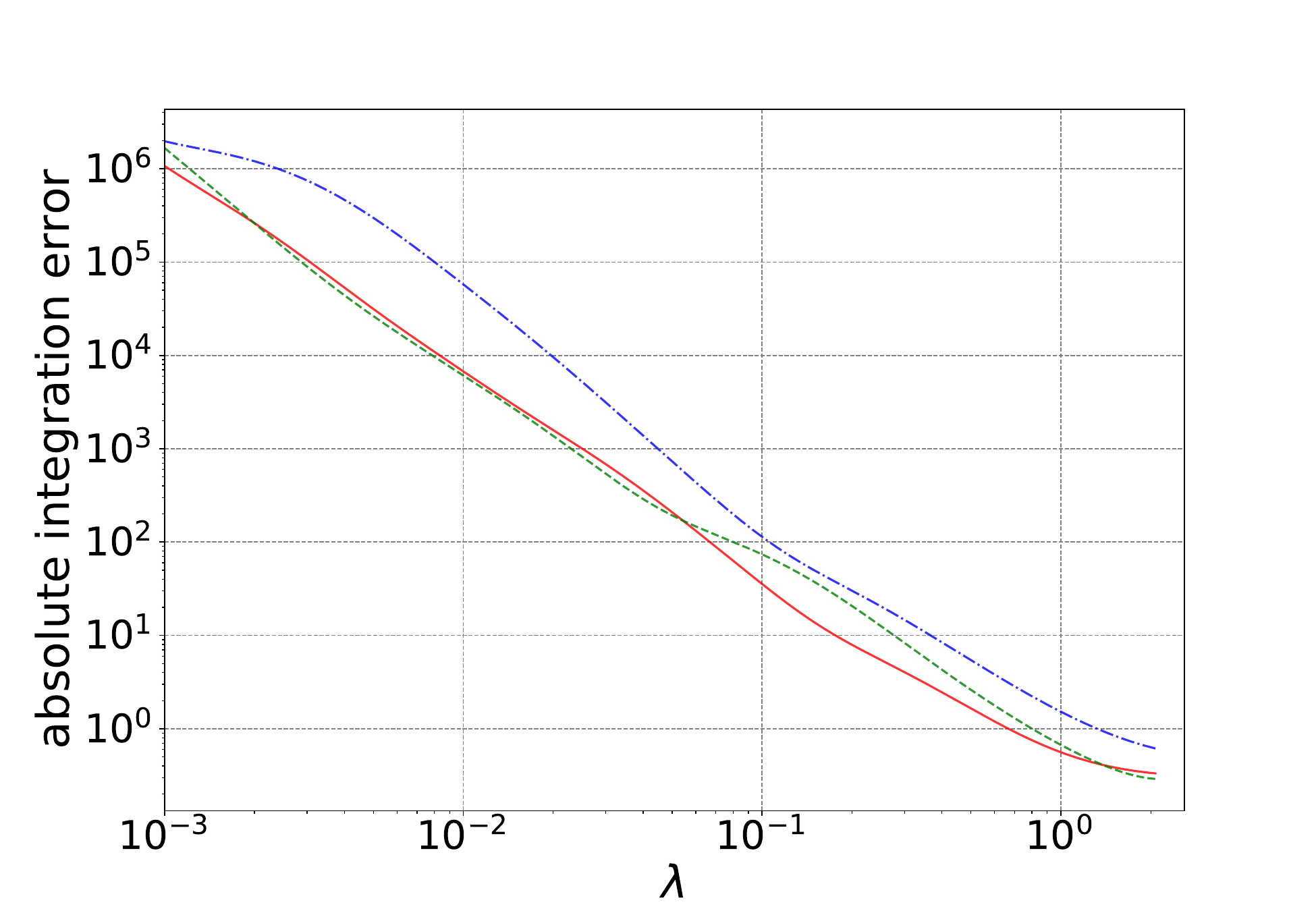}
  \includegraphics[width=0.495\textwidth]{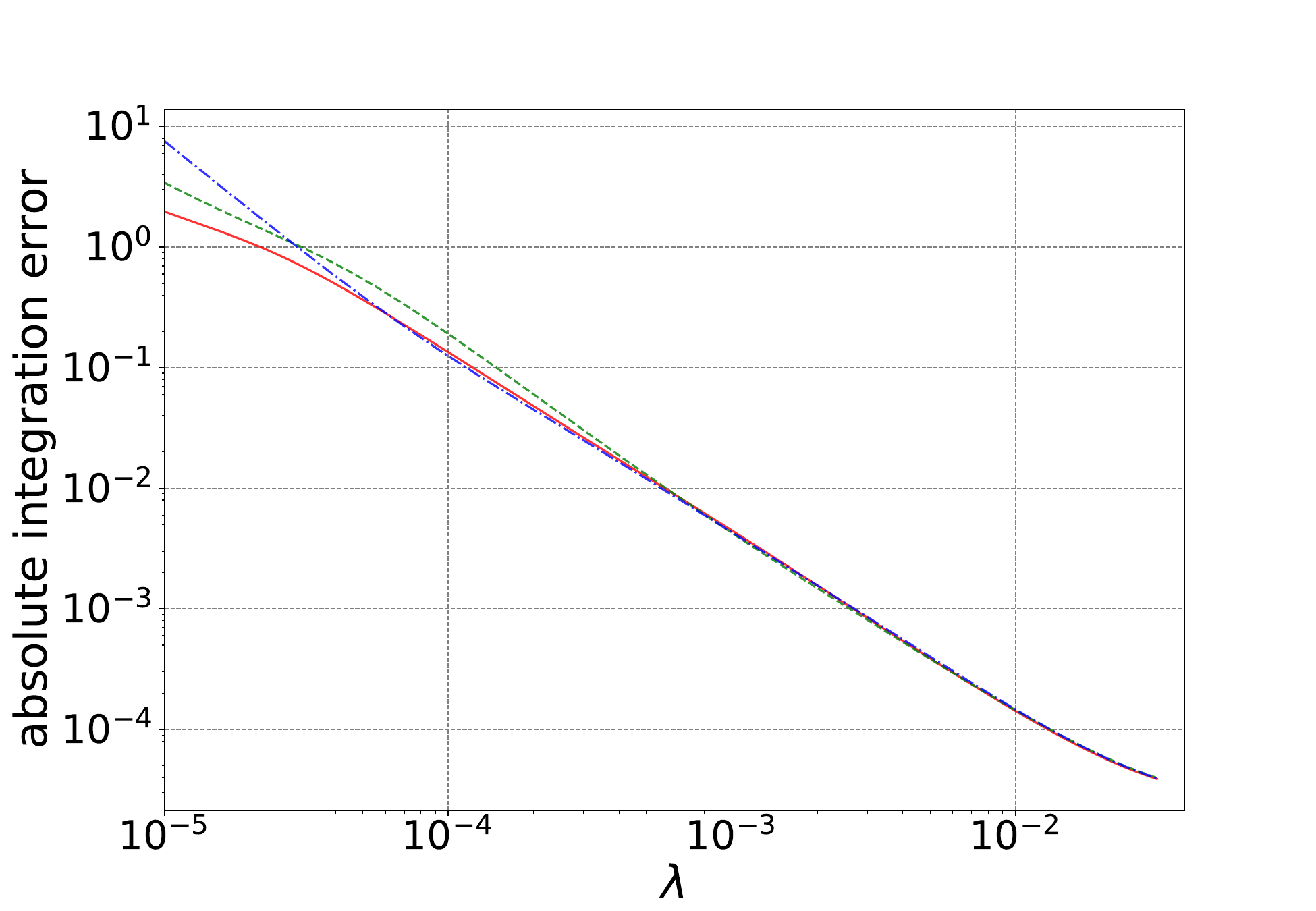}
  \caption{Absolute Monte Carlo integration errors for the first
    sector of the ladder2L (left) and the first sector of the elliptic2L
    (right) example as a function of a global $\lambda=\lambda_j$.
    For each case three different samples of $10^6$
    $\vec{x}_{(i)}$ are shown. The plots end at the largest values of
    $\lambda$ that still give valid contours.}
  \label{fig:abserr-by-1d-lambda}
\end{figure}

Numerically solving the contour integral introduced in \ref{eq:contour},
\begin{align}
  I=\int\limits_{0}^{1}  \prod\limits_{j=1}^{N}\rd x_j\,
  \det\!\left(\frac{\partial \vec{z}(\vec{x})}{\partial \vec{x}}\right)
             {\cal I}(\vec{z}(\vec{x})) \; ,
\end{align}
with the contour deformation defined in \ref{eq:deformp} still leaves the question  how to choose optimal values for $\vec{\lambda}$, and the functional form is not necessarily optimal. The Monte Carlo estimate of the integral is
\begin{align}
  I\approx I_n =\frac{1}{n}\sum_{i=1}^n
  \det\!\left(\frac{\partial \vec{z}(\vec{x}_{(i)})}{\partial \vec{x}_{(i)}}\right)
  \mathcal{I}(\vec{z}(\vec{x}_{(i)}))\; .
\end{align}
Its statistical error is minimized if the integrand approaches a constant,
\begin{align}
  \det\!\left(\frac{\partial \vec{z}(\vec{x})}{\partial \vec{x}}\right)
  \mathcal{I}(\vec{z}(\vec{x}))\approx \text{const}\;.
\label{eq:param_require}
\end{align}
Correspondingly, to construct an optimal contour through a neural network we use the variance of the Monte Carlo integration for large $n$ as
the loss function,
%
\begin{align}
  L = \sigma^2_n = \frac{1}{n-1}\sum_{i=1}^n
  \left|\,
  \det\!\left(\frac{\partial \vec{z}(\vec{x}_{(i)})}{\partial \vec{x}_{(i)}}\right) \mathcal{I}(\vec{z}(\vec{x}_{(i)})) 
  - I_n\,\right|^2,
  \label{eq:variance_loss}
\end{align}
All terms inside the absolute value squared are complex numbers. Note that the loss function has to be real valued. 

\subsection{Global complex shift}
\label{sec:global_shift} 

\begin{figure}[t]
  \includegraphics[width=0.99\textwidth]{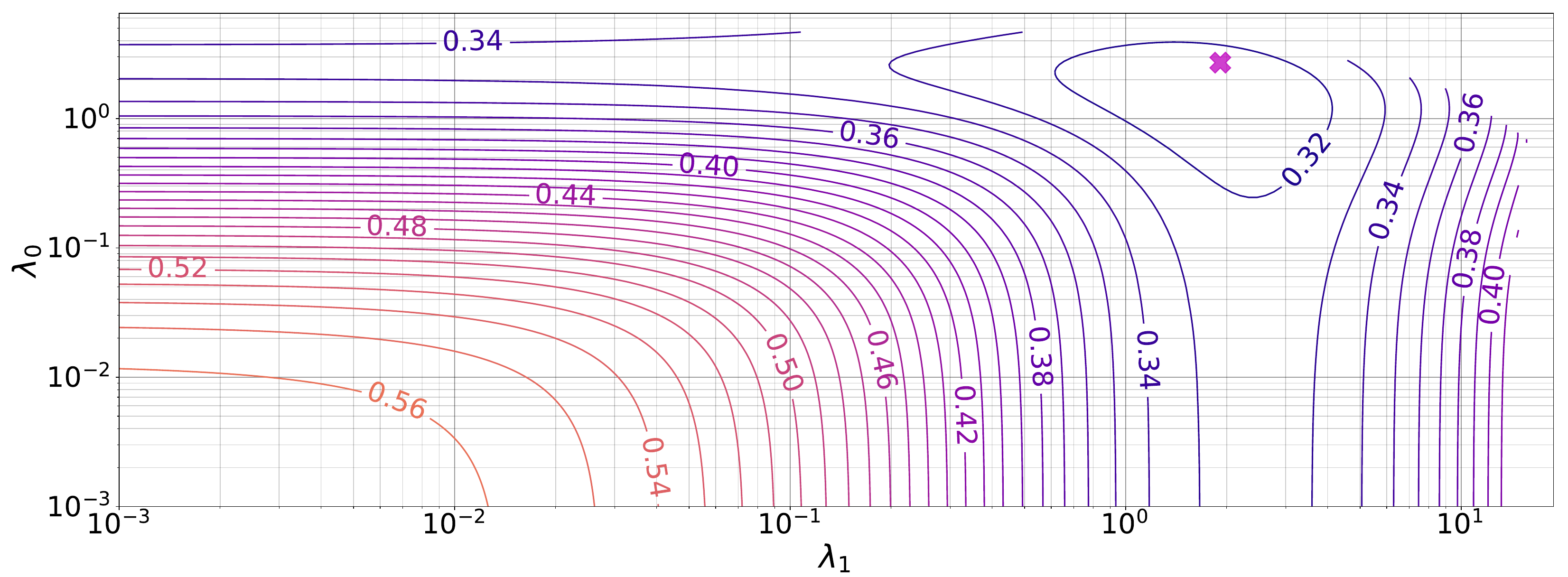}
  \caption{Absolute Monte Carlo integration error for the first sector
    of the ladder2L example, depending on $\lambda_0$ and $\lambda_1$,
    centered around $\vec{\lambda}=(2.71, 1.91, 2.73, 0.46, 5.91,
    1.59)$ (marked with a cross), which is the optimum point selected
    by $\Lambda$-glob. Note that integration error only varies between
    0.32 and 0.56 over 3 orders of magnitude in $\lambda_0$ and $\lambda_1$. 
    We find this to be a common feature.}
  \label{fig:abserr-by-2d-lambda}
\end{figure}

The standard \pysecdec{} approach of choosing the deformation
parameters works fast because it only requires to evaluate $F(\vec{x})$
and its derivatives on a set of e.g. $10^4$ points, and often produces
$\lambda_j$ that are good enough in practice. For challenging
integrals, however, it is useful to invest time into improving the
$\lambda_j$. For this purpose, we search for deformation parameters
$\lambda_j$ which minimize the Monte Carlo integration error or the
loss function $L$ on a reduced set of points. 

\begin{figure}[t]
  \includegraphics[width=0.495\textwidth]{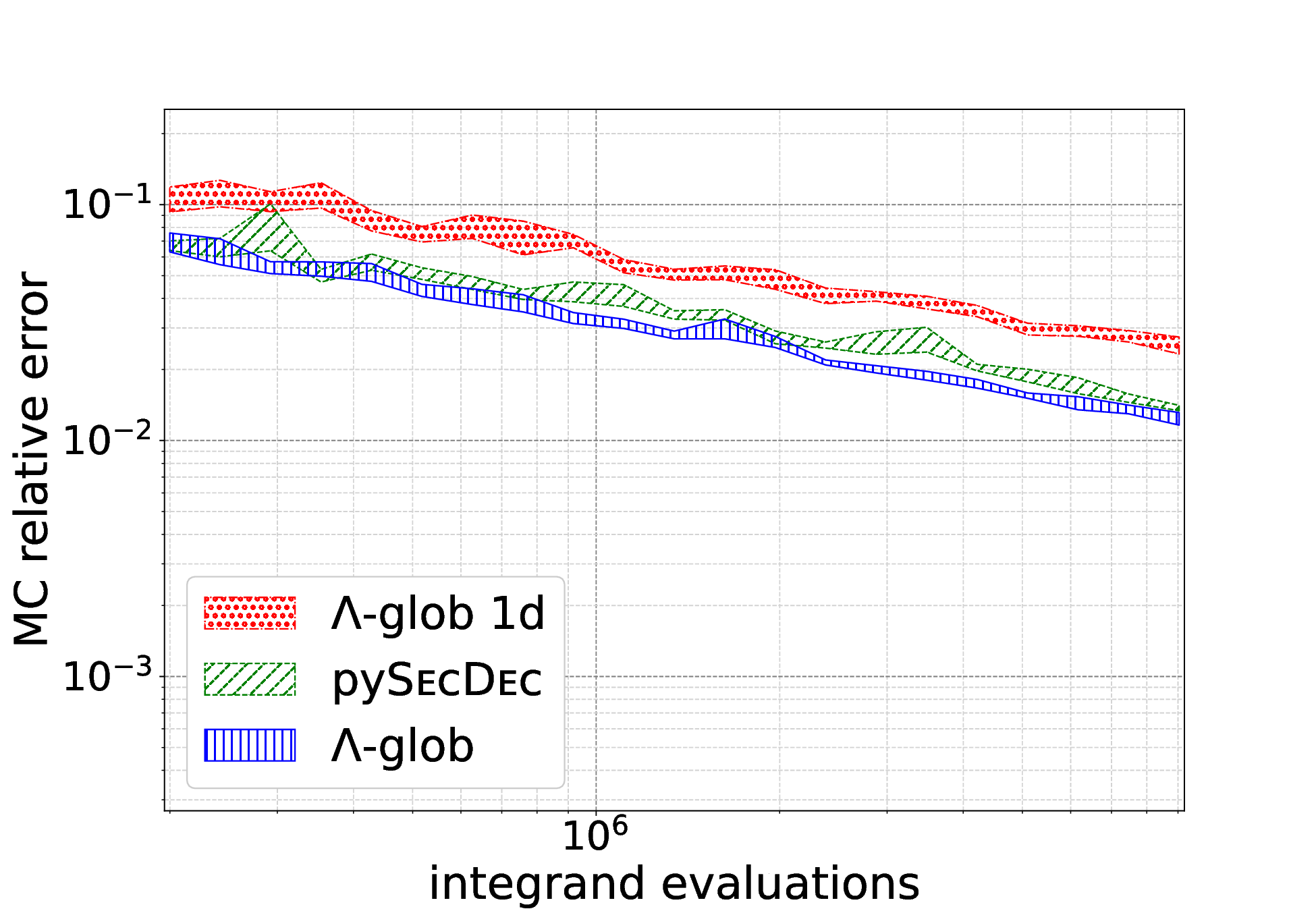}
  \includegraphics[width=0.495\textwidth]{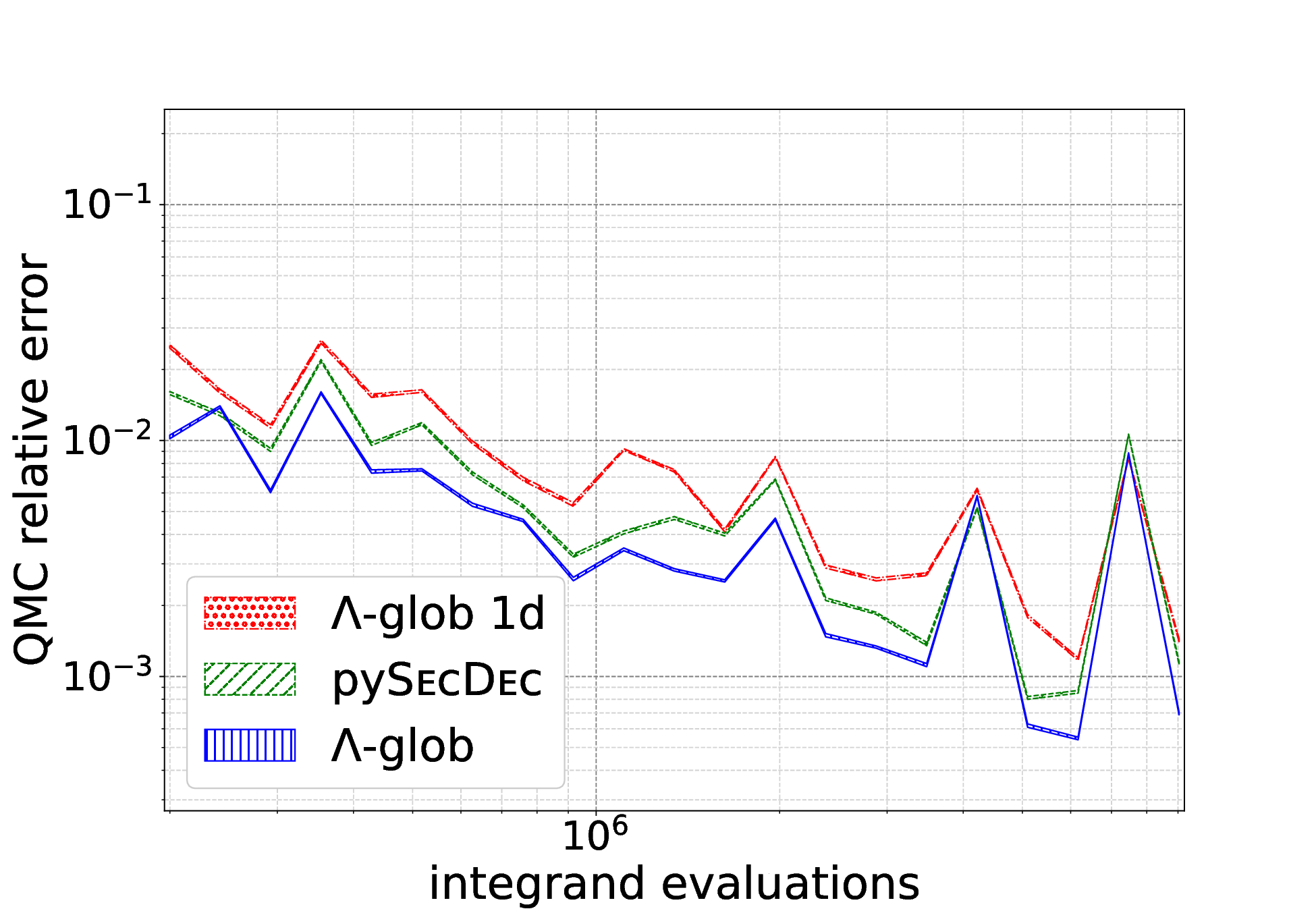}
  \includegraphics[width=0.495\textwidth]{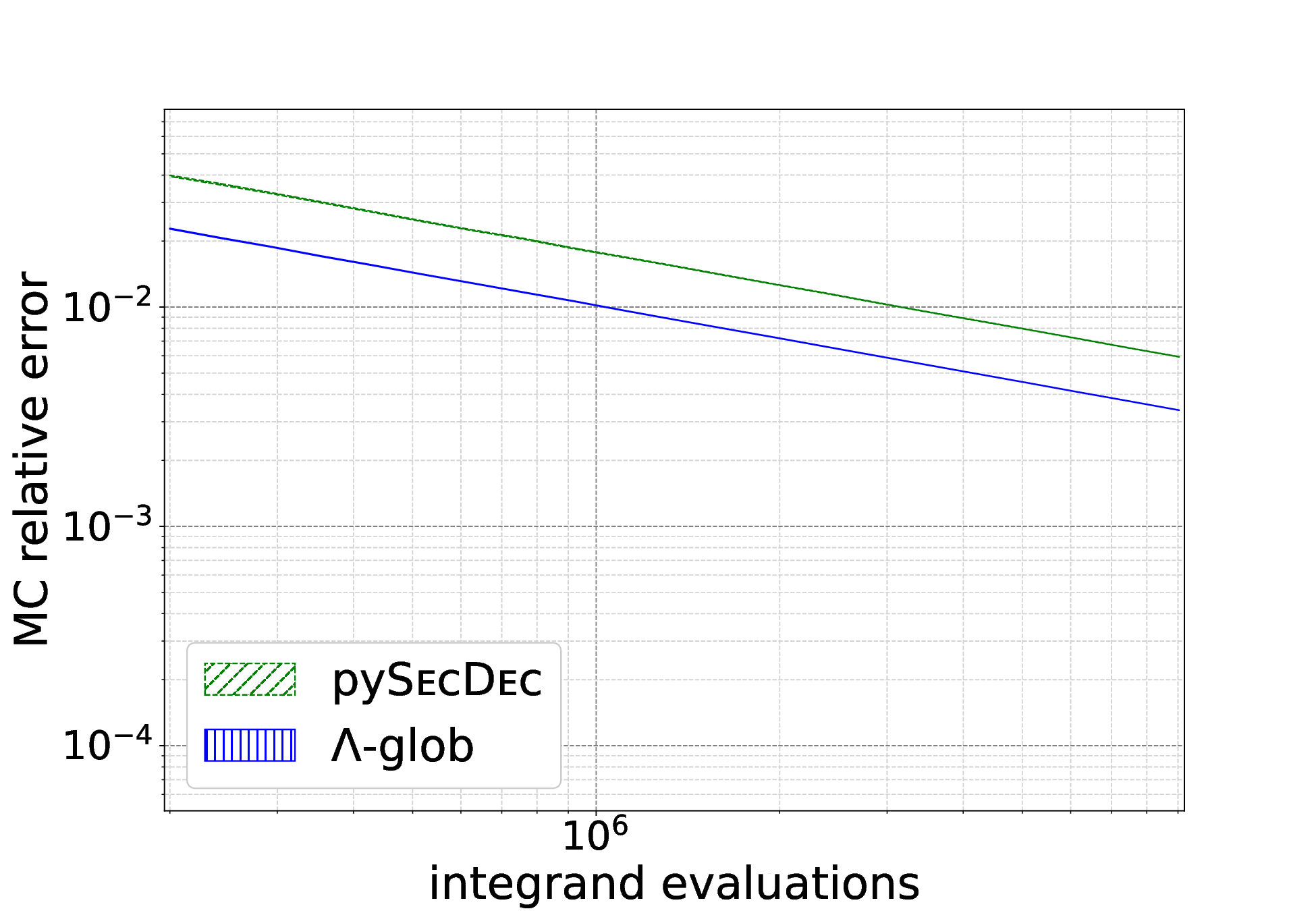}
  \includegraphics[width=0.495\textwidth]{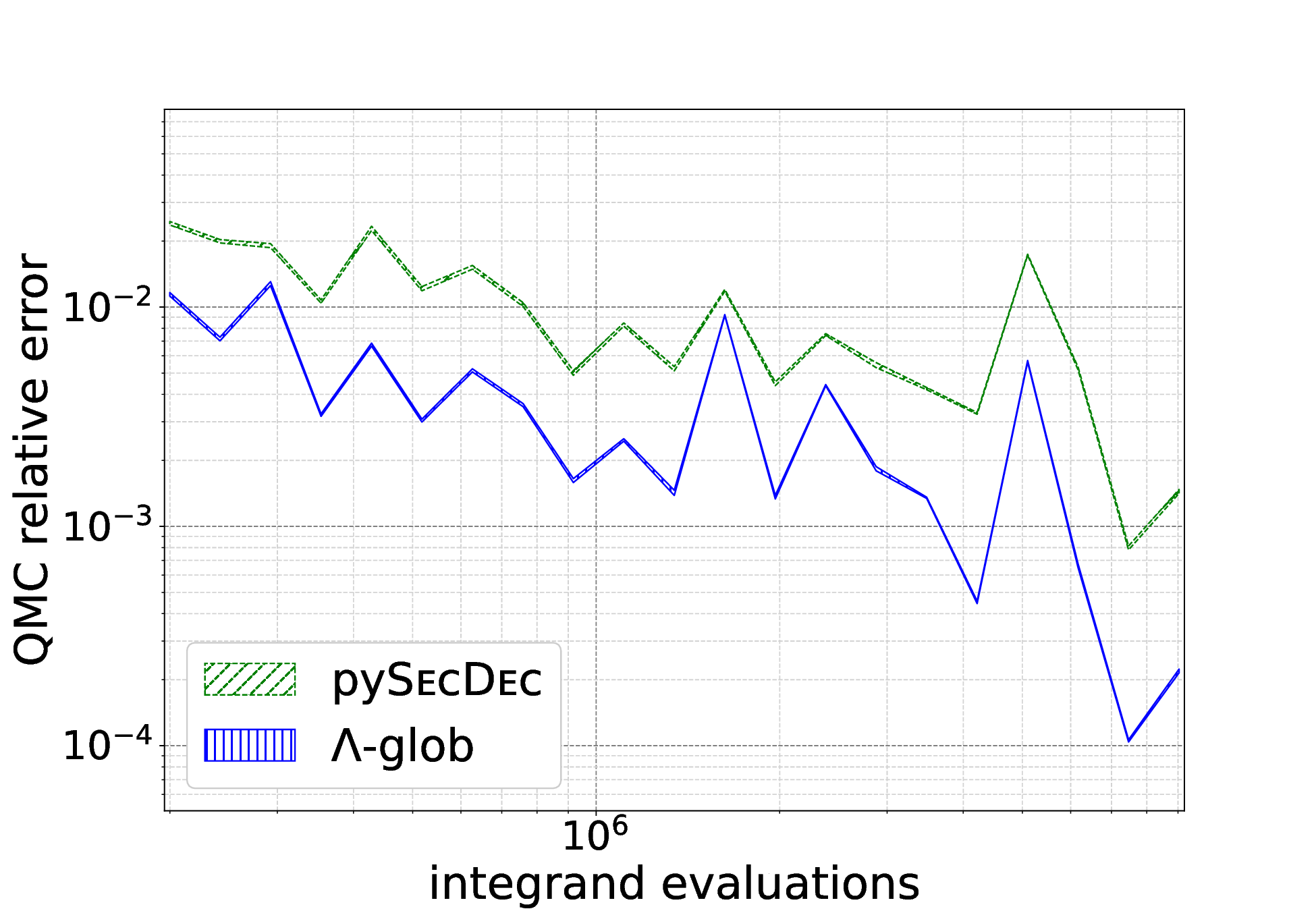}
  \caption{Monte Carlo (left) and Quasi Monte Carlo (right)
    integration errors for the first sector of the ladder2L example
    (upper) and the elliptic2L example (lower). For the
    ladder2L diagram we show the 1-dimensional and $n$-dimensional $\lambda$-selections using $\Lambda$-glob; for the elliptic2L diagram these two turn out to be the same, so only one is shown.}
  \label{fig:relerr-by-lambda-choice}
\end{figure}

The beneficial effect of tuning the $\lambda_j$ values is illustrated in
\ref{fig:abserr-by-1d-lambda} and \ref{fig:abserr-by-2d-lambda}. Two
aspects complicate this minimization problem.  First, the Monte Carlo
integration error depends on the reference sample and can be noisy, as
shown for the ladder2L example in
\ref{fig:abserr-by-1d-lambda}: different choices of the $10^6$
sampling points lead to a variation of the integration error by up to
an order of magnitude. Minimizing the loss on a frozen set will
overfit to the selected points and lead to an non-optimal choice for the
actual integration. Second, the allowed region in the $\lambda$-space has
a non-trivial shape and is currently only determined by searching for
sign check errors. This noisy determination of the allowed region
becomes a problem in practice, because the optimal $\lambda_j$ often
lie close to this boundary, as can be seen from
\ref{fig:abserr-by-1d-lambda}. As a side remark, this is why the
$\lambda$-construction in standard \pysecdec{}  can simply choose the
largest possible $\lambda_j$-vector in some predetermined direction
and still work well in practice.

\begin{algorithm}[b!]
  \caption{The $\Lambda$-glob algorithm for the global lambda optimization. Good default settings for the tested examples were $\lambda^{(0)}_j=0.1$, $s^{(0)}_j=2$, $\eta^-=0.5$, $\eta^+=1.125$, $\beta=0.1$, and $n=100$.}
  \begin{small} \begin{algorithmic}
        \REQUIRE $L(\vec{x}, \ell_j)$: The loss function (given by e.g. \ref{eq:variance_loss}) with parameters  $\lambda_j=e^{\ell_j}$
        \REQUIRE $\mathcal{E}_\text{sign}(\vec{x}, \ell_j)$: Sign check error boolean function with parameters $\lambda_j=e^{\ell_j}$
        \REQUIRE $s^{(0)}_j$: The initial step size vector
	    \REQUIRE $0<\eta^-<1<\eta^+$: Step size decrease and increase factors
	    \REQUIRE $\beta>0$: The size of the backtracking step
        \REQUIRE $\lambda^{(0)}_j$: Initial parameter vector
        \REQUIRE $n$: The number of the optimization iterations
         \STATE $\ell_j^{(0)}\leftarrow\log\lambda^{(0)}_j$ (Initialize log parameter)
    	\FOR{$t=1,\dots,n$}
    	\STATE Draw a sample $\vec{x}^{(t)}$ from a unit hypercube
    	\STATE $\widehat{\ell}_j \leftarrow \ell^{(t-1)}_j$
    	\WHILE{$\mathcal{E}_\text{sign}(\vec{x}^{(t)}, \widehat{\ell}_j)$}
    	\STATE $\widehat{\ell}_j\leftarrow \widehat{\ell}_j - \beta$
    	\ENDWHILE
        \STATE $\widehat{\ell}_j\leftarrow \widehat{\ell}_j - s^{(t-1)}_j\,\text{sgn}\!\left(\frac{\partial L}{\partial \ell_j}\!\left(\vec{x}^{(t)}, \widehat{\ell}_j\right)\right)$
        \IF{$L\!\left(\vec{x}^{(t)}, \widehat{\ell}_j\right)>L\!\left(\vec{x}^{(t)}, \ell^{(t-1)}_j\right)$}
        \STATE $\ell^{(t)}_j \leftarrow \ell^{(t-1)}_j$
    	\STATE $s_j^{(t)}\leftarrow\eta^- s_j^{(t-1)}$
    	\ELSE
    	\STATE $\ell^{(t)}_j \leftarrow \widehat{\ell}_j$
    	\STATE $s_j^{(t)} \leftarrow \left(\mathrm{if} \left(\frac{\partial L}{\partial \widehat{\ell}_j}\right)\cdot\left(\frac{\partial L}{\partial \ell^{(t-1)}_j}\right)>0 \mathrm{\ then\ } \eta^+ \mathrm{\ else\ } \eta^-\right)s_j^{(t-1)}$
    	\ENDIF
    	\ENDFOR
    	\STATE $\lambda_j\leftarrow\exp\!\left(\ell^{(n)}_j\right)$
    	\RETURN $\lambda_j$
	\end{algorithmic} \end{small}
	\label{al:global_lambda}
\end{algorithm}

To circumvent these problems in optimizing the $\lambda_j$ we
introduce the $\Lambda$-glob algorithm, a modified version of the
Rprop algorithm~\cite{RB93,IH2000} with the added explicit handling of the allowed
region for $\lambda_j$. A detailed description is given in \ref{al:global_lambda}.
We start with a point $\lambda^{(0)}_j$ and a step size
$s^{(0)}_j$. To converge faster to a potentially far-away minimum we
work with a logarithmic scale $\ell_j=\log \lambda_j$. In the first
step, called \textit{backtracking}, the algorithm decreases $\ell_j$
by some increment $\beta$ if a sign check error
$\mathcal{E}_\text{sign}(\ell_j)$ occurs. After backtracking, we
choose a proposal point $\hat{\ell}_j$ employing gradient decent. If
the loss for the proposal point is smaller than for the previous point
it is kept and the step size $s_j$ is decreased (increased) depending
on whether the gradient of the loss has changed (not changed) its
sign. If the loss is larger, the proposal point is rejected and the
step size $s_j$ is decreased by a factor of~$\eta^-$. All these steps are
repeated a predetermined number of times. To avoid overfitting, we draw a new sample of points $\vec{x}_{(i)}$ for each iteration.

The $\Lambda$-glob algorithm benefits from a large initial step size that allows to efficiently step over local minima of $L$. Because the step size is adjusted automatically we do not have to define a learning schedule, unlike for standard gradient descent
optimization.

In \ref{fig:abserr-by-2d-lambda} we show the landscape of the loss
function and observe that our algorithm has found the global minimum and
that this minimum is very broad. While it is very flat in $\lambda_0$
and $\lambda_1$ individually, a correlated shift increases the loss
function more steeply.  Results of this algorithm for different
Feynman integrals are presented in
\ref{fig:relerr-by-lambda-choice}. In addition to a standard Monte Carlo algorithm we also show the result for the Quasi Monte Carlo algorithm in \pysecdec{}\cite{Borowka:2018goh}. Depending on the integral and the
kinematic configuration $\Lambda$-glob gives comparable or improved results
compared to the standard \pysecdec{} construction.

\subsection{Generalized local transformation}
\label{sec:local_trafo} 

Moving beyond the optimization of a global deformation parameter, we can exploit the full freedom of the reparametrization with a local transformation of $\lambda$ to further minimize the Monte Carlo error. In principle any reparametrization in \ref{eq:deformp} would serve our purpose as long as the defined contour does not cross or enclose any singularities. However, a good contour should also obey the following criteria:
\begin{enumerate}
    \item The Monte Carlo error should be minimized, \ie the product of the Jacobian of the transformation and the integrand should be nearly constant, see \ref{eq:param_require};
    \item The endpoints have to be fixed for Cauchy's theorem to be applicable as in \ref{eq:contour};
    \item The parametrization should be numerically stable and, if possible, have tractable Jacobians. A tractable Jacobian is not only more stable numerically, it also helps to make sure that the procedure gives meaningful results.
\end{enumerate}
Let us consider again the integral along a contour $\gamma$,
\begin{align}
    I = \int_\gamma \prod\limits_{j=1}^{N}\rd z_j \,{\cal I}(\vec{z}) \; .
\end{align}
We parametrize this contour in terms of the real parameters $y_j\in[0, 1]$,
%
\begin{align}
    z_j &= y_j - \mathrm{i} \tau_j(\vec{y}) 
    \qquad \text{and} \qquad 
    \tau_j = \lambda_j(\vec{y})\,y_j(1-y_j)\frac{\partial F(\vec{y})}{\partial y_j} \; ,
    \label{eq:gen_lambda}
\end{align}
where the form of the imaginary part guarantees the correct boundary conditions. In contrast to \ref{eq:deformp}, the deformation $\lambda_j$ is now a \textit{local} parameter, depending on $\vec{y}$. 

To minimize the variance in the numerical integration, $\vec{y}$ needs to be sampled according to some non-trivial probability distribution. In practice, this probability distribution is neither known nor is it possible to easily sample from it. Therefore, we introduce an additional mapping
\begin{align}
y_j=f_j(\vec{x})\;,
\label{eq:second_trafo}
\end{align}
with uniformly distributed $x_j\in[0,1]$. With all these transformations the integral becomes
\begin{align}
\begin{split}
\int_\gamma \prod\limits_{j=1}^{N}\rd z_j\,\mathcal{I}(\vec{z})&= \int\limits_{0}^{1} \prod\limits_{j=1}^{N}\rd y_j\,\det\!\left(\frac{\partial \vec{z}(\vec{y})}{\partial \vec{y}}\right)\mathcal{I}(\vec{z}(\vec{y}))\\
&= \int\limits_{0}^{1}\prod\limits_{j=1}^{N}\rd x_j\,\det\!\left(\frac{\partial \vec{z}(\vec{y})}{\partial \vec{y}}\right) \det\!\left(\frac{\partial \vec{y}(\vec{x})}{\partial \vec{x}}\right)\mathcal{I}(\vec{z}(\vec{y}(\vec{x})))\;.
\end{split}
\end{align}
Except for the boundaries, the functions $\lambda$ and $f$ can be chosen freely. A flexible and promising way to parametrize these functions is with neural networks. 
A critical aspect of the reparametrization are the Jacobians
\begin{align}
\left(J_\lambda\right)_{jk}=\frac{\partial z_j}{\partial y_k} =\frac{\partial (y_j-\imag\tau_j)}{\partial y_k}
=\delta_{jk}-\imag\frac{\partial\tau_j}{\partial y_k}
\qquad \text{and} \qquad 
\left(J_f\right)_{jk}=\frac{\partial y_j}{\partial x_k} =\frac{\partial f_j(\vec{x})}{\partial x_k}\; , 
\end{align}
the first of which is complex. For these Jacobians to be non-singular, we require our mappings to be bijective. While the complex Jacobian is always non-singular by construction, we have to ensure explicitly that the function $f$ is bijective. The function $\lambda$ does not have to be bijective. However, one needs to ensure that the sub-Jacobian
\begin{align}
\begin{split}
\frac{\partial\tau_j}{\partial y_k} =& \, \frac{\partial\lambda_j(\vec{y})}{\partial y_k}\,y_j(1-y_j)\,\frac{\partial F(\vec{y})}{\partial y_j}+\\
&+\lambda_j(\vec{y})\,\delta_{jk}(1-2y_j)\,\frac{\partial F(\vec{y})}{\partial y_j}
+\lambda_j(\vec{y})\,y_j(1-y_j)\,\frac{\partial^2 F(\vec{y})}{\partial y_j \partial y_k}
\end{split}
\end{align}
is numerically stable.

\subsection{Normalizing flow setup}
\label{sec:normalizing_flow}

The reasons to split the full mapping  $\vec{x}\to\vec{z}$ into the real mapping $\vec{x}\to\vec{y}$ and the complex mapping $\vec{y}\to\vec{z}$ as in \ref{eq:gen_lambda} and \ref{eq:second_trafo} are the following: First, it will allow us to use a normalizing flow~\cite{nflow1,Kobyzev_2020,papamakarios2019normalizing,nflow_review,mller2018neural} for the real mapping, giving us a  tractable Jacobian. The Jacobian of the complex mapping needs to be evaluated numerically and  is computationally more expensive than the Jacobian of the normalizing flow. 
Second, when we evaluate a kinematic phase-space point which does not require any contour deformation, as there are no integrable singularities, we just turn off the complex mapping. The real mapping then becomes a version of neural importance sampling~\cite{maxim,Bothmann:2020ywa,Gao:2020vdv,Gao:2020zvv,Chen:2020nfb,Danziger:2021eeg}.

To train our network we will use the variance loss defined in \ref{eq:variance_loss}. From the discussion of the $\Lambda$-glob algorithm and \ref{fig:abserr-by-1d-lambda} we know that the integration error is fairly insensitive to small changes in $\lambda$. Furthermore, for the parametrization in \ref{eq:gen_lambda}, we found that making $\lambda$ local (i.e. dependent on $\vec{x}$) also hardly affects the loss. Because the introduction of a neural network comes with a computational cost, we keep $\lambda$ global in our NN-approach. This means we rely on the $\Lambda$-glob algorithm to first find optimized $\lambda_j$ and then use a normalizing-flow network 
to optimize the sampling of the real parameters and minimize the variance. 
In our experiments we perform the numerical loop integration for various Feynman diagrams given in \ref{fig:feyn_diags}, which are represented in the $N$-dimensional Feynman parameter space.

\subsubsection*{Network architecture}

Normalizing flows encode a bijective mapping between a physics and a latent space. The model can be evaluated in either direction with comparable
efficiencies, at least in the invertible network (INN)
variant~\cite{inn,coupling2,glow}. Even if we are not interested in
this symmetric evaluation, normalizing flows have the considerable
advantage of a tractable Jacobian. A simple realization are stacked coupling layers~\cite{coupling1,coupling2}, where we split the
input vector $x$ in $x_1$ and $x_2$ and use an element-wise
multiplication~$\odot$ and sum to define the mapping
%
\begin{align}
y_1 &= x_1 \odot e^{s_1(x_2)} + t_1(x_2)
\qquad & 
x_1 &= \left( y_1 - t_1(x_2) \right) \odot e^{-s_1(x_2)}\notag  \\
y_2 &= x_2 \odot e^{s_2(y_1)} + t_2(y_1)
\qquad & 
x_2 &= \left( y_2 - t_2(y_1) \right) \odot e^{-s_2(y_1)} \; ,
\label{eq:CouplingEx}
\end{align}
where $s_1, s_2, t_1$ and $t_2$ are parametrized by neural networks. 
The Jacobian of such a coupling block is~\cite{coupling2} 
\begin{align}
J 
&= \begin{pmatrix} \mathbbm{1} & 0 \\ \frac{\partial y_2 }{\partial y_1} & \diag(e^{s_{2}(y_1)}) \end{pmatrix} 
\begin{pmatrix} \diag(e^{s_{1}(x_2)}) & \frac{\partial y_1 }{\partial x_2} \\ 0 &  \mathbbm{1} \end{pmatrix} \; .
\end{align}
%
While $J$ is not triangular, we will only be interested in the log-determinant, which can be calculated efficiently as
\begin{align}
\begin{split}
\log\left(\det J\right) 
&=\log\!\left(\prod\limits_{i=1}^{\dim x_2} e^{s_1(x_2)_i}\right) +
\log\!\left(\prod\limits_{i=1}^{\dim y_1} e^{s_2(y_1)_i}\right)\\
&= \sum_{i=1}^{\dim x_2} s_1(x_2)_i + \sum_{i=1}^{\dim y_1} s_2(y_1)_i \; .
\end{split}
\label{eq:inn_determinant}
\end{align}

\begin{figure}[t]
  \centering
  \includegraphics[page=1,width=0.99\textwidth]{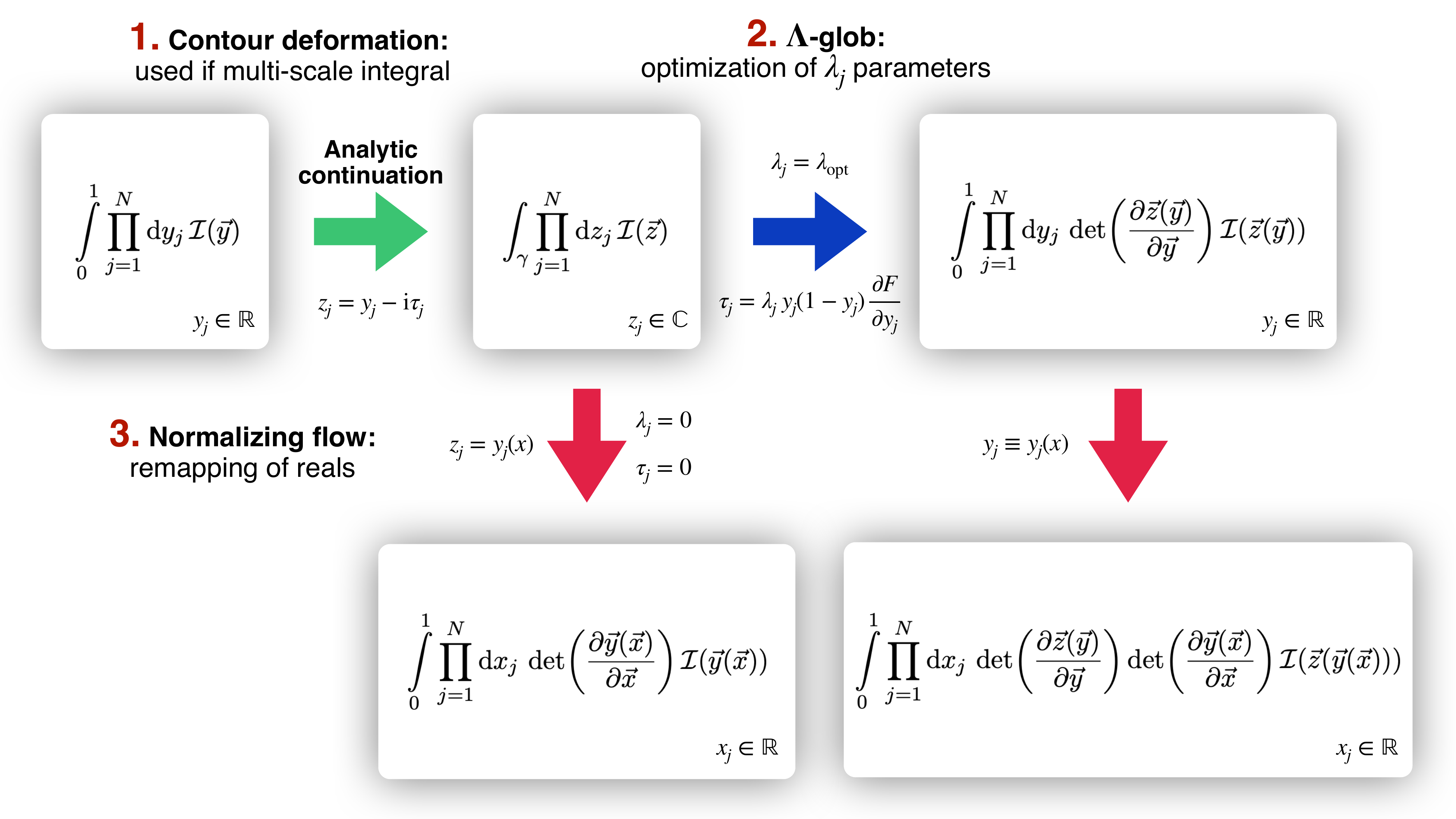}
  \caption{Schematic illustration of our workflow.}
  \label{fig:workflow}
\end{figure}

For all examples we employ a normalizing flow consisting of these affine coupling blocks, where each coupling block describes a bijective mapping $\mathbb{R}^N \leftrightarrow \mathbb{R}^N$. To map the Feynman parameters $x\in[0,1]^N$ from the unit-hypercube to $\mathbb{R}^N$ bijectively we apply the logit function
\begin{align}
    y = \mathrm{logit}(x) \equiv \log\!\left(\frac{x}{1-x}\right),
\qquad \text{with} \qquad 
\left(J_\text{logit}\right)_{jk} = \frac{\delta_{jk}}{x_j-x_j^2} \; ,
\end{align}
which is the inverse of the sigmoid function
\begin{align}
    y = \mathrm{sig}(x)\equiv\frac{1}{1 + \exp(-x)} \; ,
    \qquad \text{with} \qquad
    \left(J_\text{sig}\right)_{jk}= \delta_{jk}\,\mathrm{sig}(x_j)(1-\mathrm{sig}(x_j))\; .
\label{eq:Jacobian_sigmoid}
\end{align}
As both Jacobians are diagonal these functions can be easily combined with the coupling blocks. For convenience, we use the sigmoid as the final network layer, such that the output domain is again the unit-hypercube. We sandwich 14 coupling blocks between the logit and sigmoid functions. In each coupling block we use a simple fully connected neural network consisting of 3 layers with 128 units and Leaky ReLU as activation function. To regularize the exponentials of the affine coupling block we use soft clamping~\cite{coupling1}, $s_\text{clamp}= c\cdot\tanh(s)$, with $c=0.5$, and activation normalization~\cite{glow}. Furthermore, we use random orthogonal matrices~\cite{cinn} to allow for more interaction between the two parts $y_1, y_2$ in the coupling blocks. Our network is implemented using \textsc{TensorFlow}~\cite{tensorflow}.

\subsubsection*{Training}

\begin{figure}[b!]
  \includegraphics[page=1,width=0.495\textwidth]{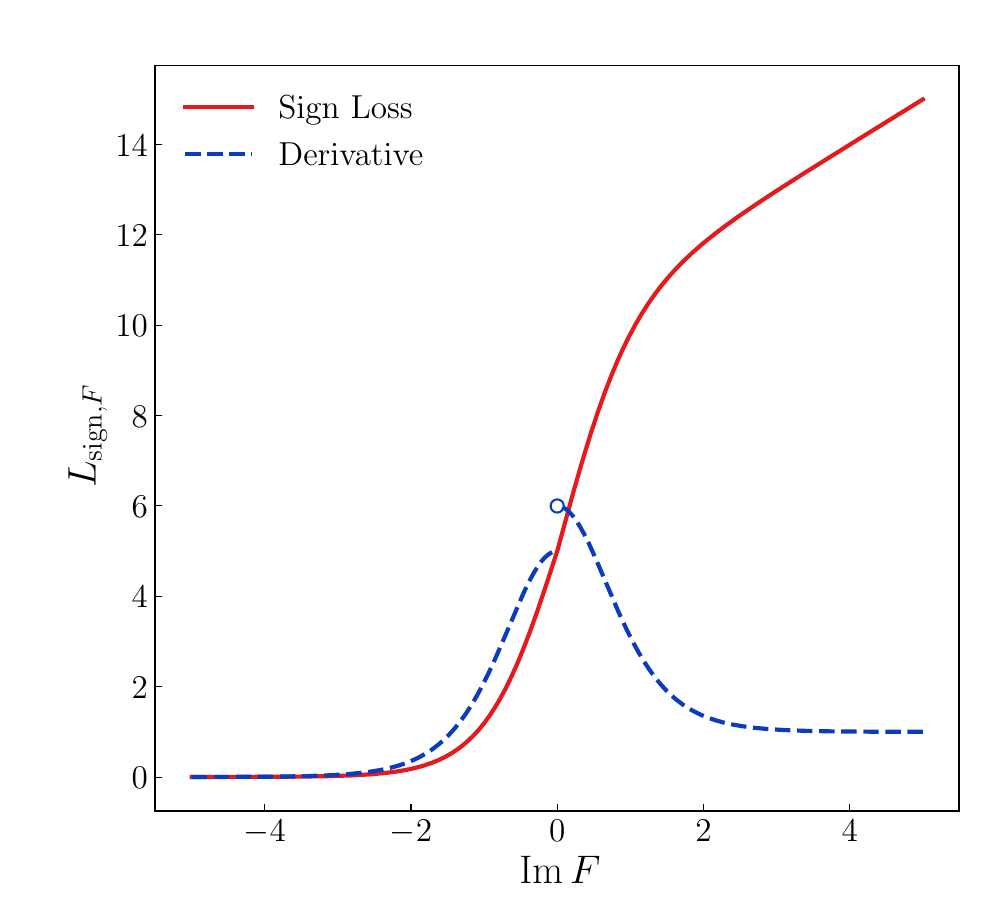}
  \includegraphics[page=1,width=0.495\textwidth]{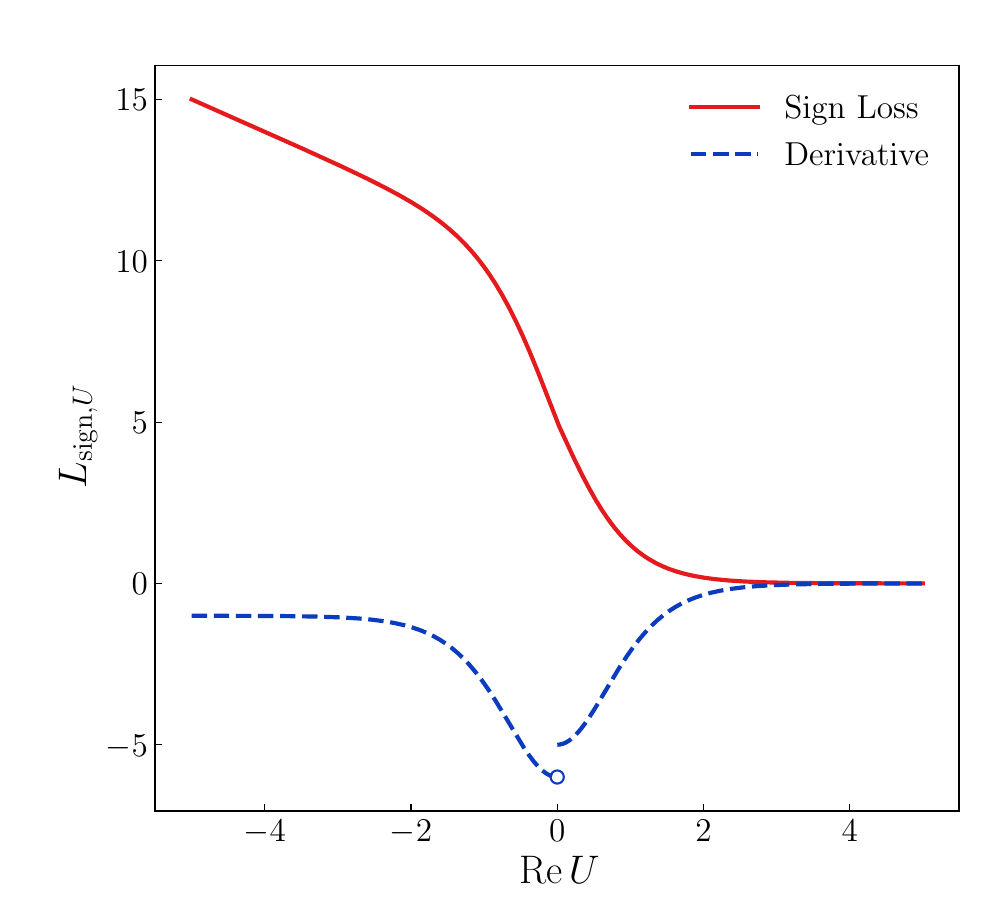}
  \caption{Illustration of the sign loss and its derivative for the $F$ (left) and $U$ (right) part for $Y=10$ and $X_{\{F,U\}}=1/2$.}
  \label{fig:sign-loss}
\end{figure}

Before introducing the neural network, we employ the $\Lambda$-glob algorithm to find optimal values of $\lambda_j$, which minimize the variance loss in \ref{eq:variance_loss} and define a valid contour on the physical Riemann sheet. Next, we train the normalizing flow to re-sample the real parameters in the spirit of neural importance sampling. This gives us a complex mapping $\vec{y}\to\vec{z}$ parametrized as in \ref{eq:deformp} with optimized $\lambda_j$, and a real mapping $\vec{x}\to\vec{y}=f(\vec{x})$ where $f$ is represented by a normalizing flow.

For kinematic phase-space regions below threshold, no contour deformation is needed. Here the $\Lambda$-glob algorithm will find $\lambda_j=0$, the complex mapping \ref{eq:deformp} will be omitted, and the real mapping alone will improve the calculation. The complete workflow is summarized in \ref{fig:workflow}.

In contrast, for kinematic phase-space points above threshold the contour is vital and we need to make sure to have the correct sign for the imaginary part of $F$ as well as for the real part of $U$. For the $\Lambda$-glob algorithm we use a simple backtracking method to discard a proposal state and step back, \ie reduce the value of $\lambda_j$, if a sign check-error occurs. For the network we add a term to the loss function. As we employ the \textsc{Adam} optimizer~\cite{adam}, this sign loss has to be differentiable, so we add
\begin{align}
    \begin{split}
    L_\text{sign}=\,Y\, \mathrm{sig}\!\left(\frac{\mathrm{Im}\,F}{X_F}\right)+\mathrm{ReLU}(\mathrm{Im}\,F)
    +\,Y\, \mathrm{sig}\!\left(-\frac{\mathrm{Re}\,U}{X_U}\right)+\mathrm{ReLU}(-\mathrm{Re}\,U)
    \end{split}
\end{align}
with $\mathrm{ReLU}(x)=\max\{0,x\}$ to the variance loss of \ref{eq:variance_loss}. Using a validation set $\vec{x}_{\text{val}}$, the relative scales $Y,X_F,X_U$ are estimated in the beginning of the training and updated every $K^\text{th}$ iteration according to
\begin{align}
    Y &= 10\cdot L_\mathrm{var}(\vec{x}_{\text{val}})\,,
    & X_F &=\frac{1}{5n}\left\vert\sum_i^n\mathrm{Im}\,F\!\left(\vec{x}_{\text{val},(i)}\right)\right\vert
    \,,
    & X_U &=\frac{1}{5n}\left\vert\sum_i^n\mathrm{Re}\,U\!\left(\vec{x}_{\text{val},(i)}\right)\right\vert\,.
\end{align}
In practice, we find that $K=10$ works well in our experiments. An illustration of the sign loss and its derivative of the $F$ and $U$ part is shown in the left and right panels of \ref{fig:sign-loss}, respectively.

\begin{figure}[t]
  \includegraphics[page=1,width=0.495\textwidth]{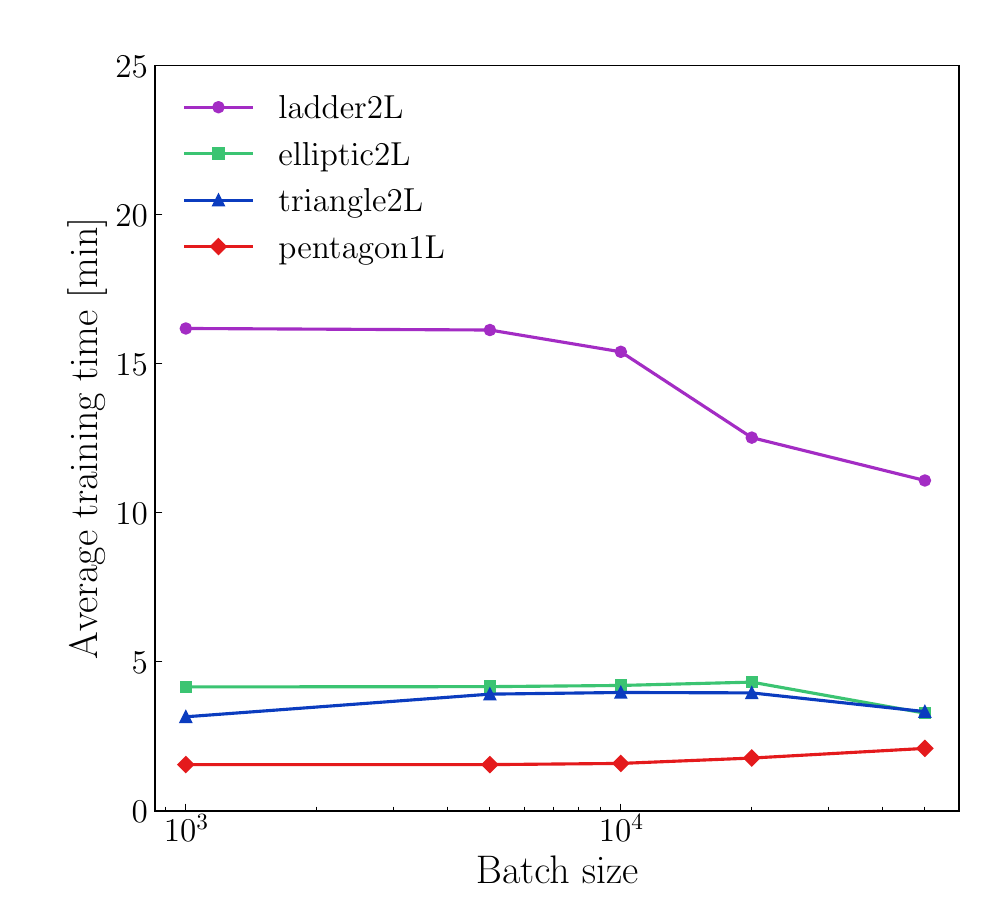}
  \includegraphics[page=1,width=0.495\textwidth]{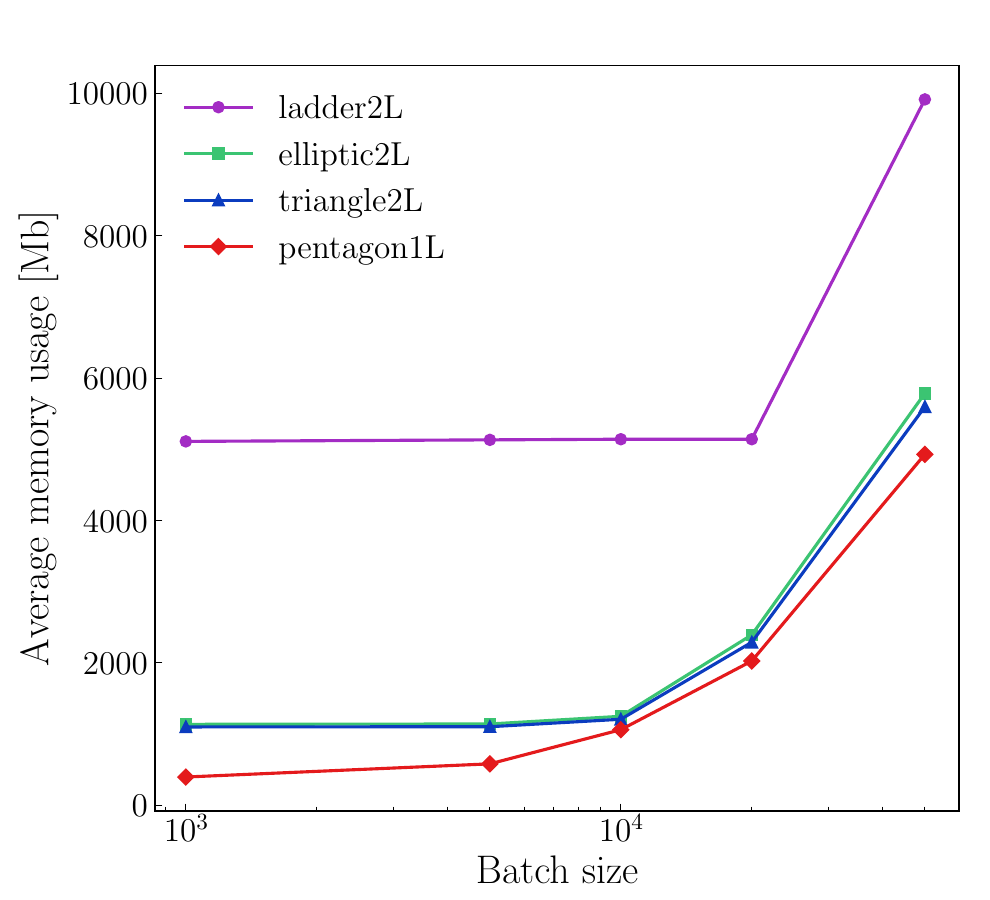}
  \caption{Average training times (left) and average memory consumption (right) for a single phase-space point as a function of the batch size $b$.}
  \label{fig:computational_costs}
\end{figure}

Moreover, a numerical bottleneck in our contour optimization is the calculation of the complex-valued determinant and its derivative. As the \textsc{TensorFlow} implementation of complex-valued determinants yields wrong gradients\footnote{See the issue raised at \textsc{GitHub}: \href{https://github.com/tensorflow/tensorflow/issues/49946}{https://github.com/tensorflow/tensorflow/issues/49946}.}, we implement our own version of the determinant. It relies on the recursive Laplace expansion and becomes computationally expensive for higher dimensional cases. This can be seen in the GPU-memory usage in \ref{fig:computational_costs}, which is significantly higher for processes involving more Feynman parameters, such as the ladder2L example.
This is one of the reasons why the timings are not competitive with the timings for the standard contour deformation in  \pysecdec{}. The largest benefit from the ML-approach is expected for high-dimensional multi-scale cases, where the contour avoiding all poles and branch cuts is a highly non-trivial hypersurface in the complex integration space. In such cases the gain in numerical precision can be so large that it outweighs the time spent to train the network. Indeed, the true advantage would show up in calculations of complete amplitudes, rather than individual integrals, containing a few integrals that would barely converge at all in \pysecdec{} but would converge well with an optimized contour.

\subsubsection*{Performance}

Finally, we illustrate the performance gain achieved by applying both, the $\Lambda$-glob algorithm only and its combination with the normalizing flow. 

\begin{figure}[t]
  \includegraphics[page=1,width=0.495\textwidth]{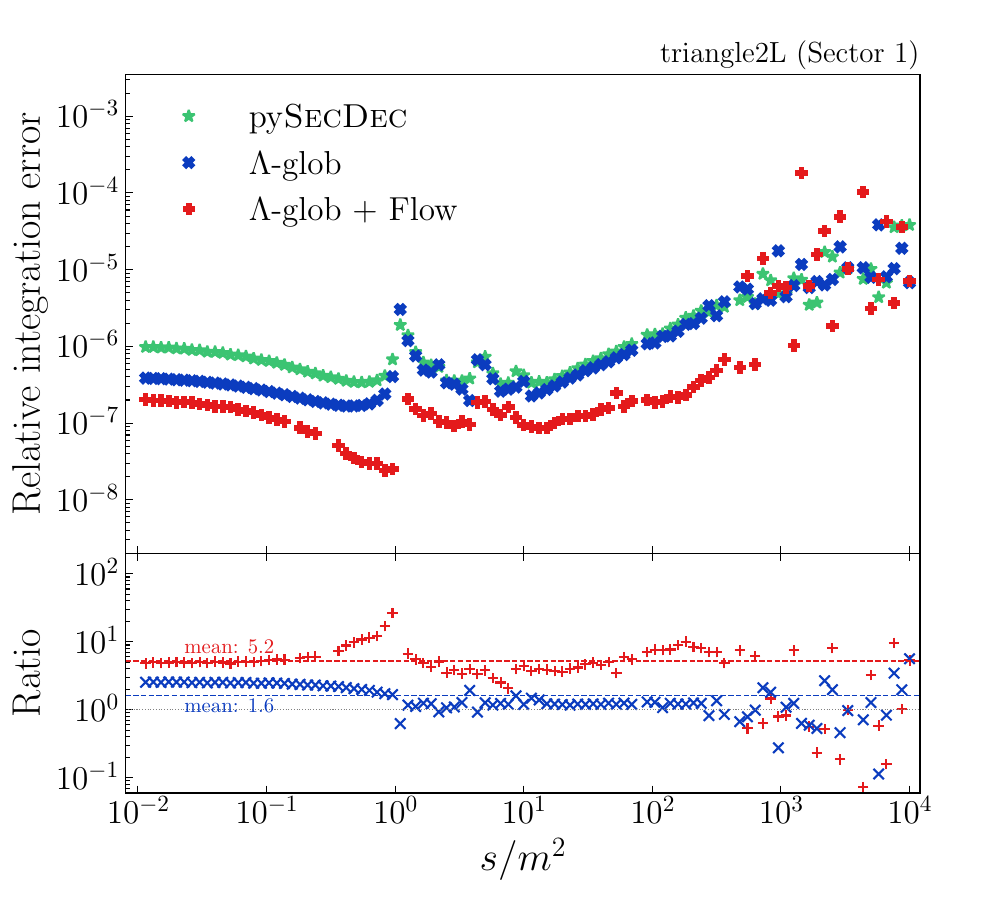}
  \includegraphics[page=1,width=0.495\textwidth]{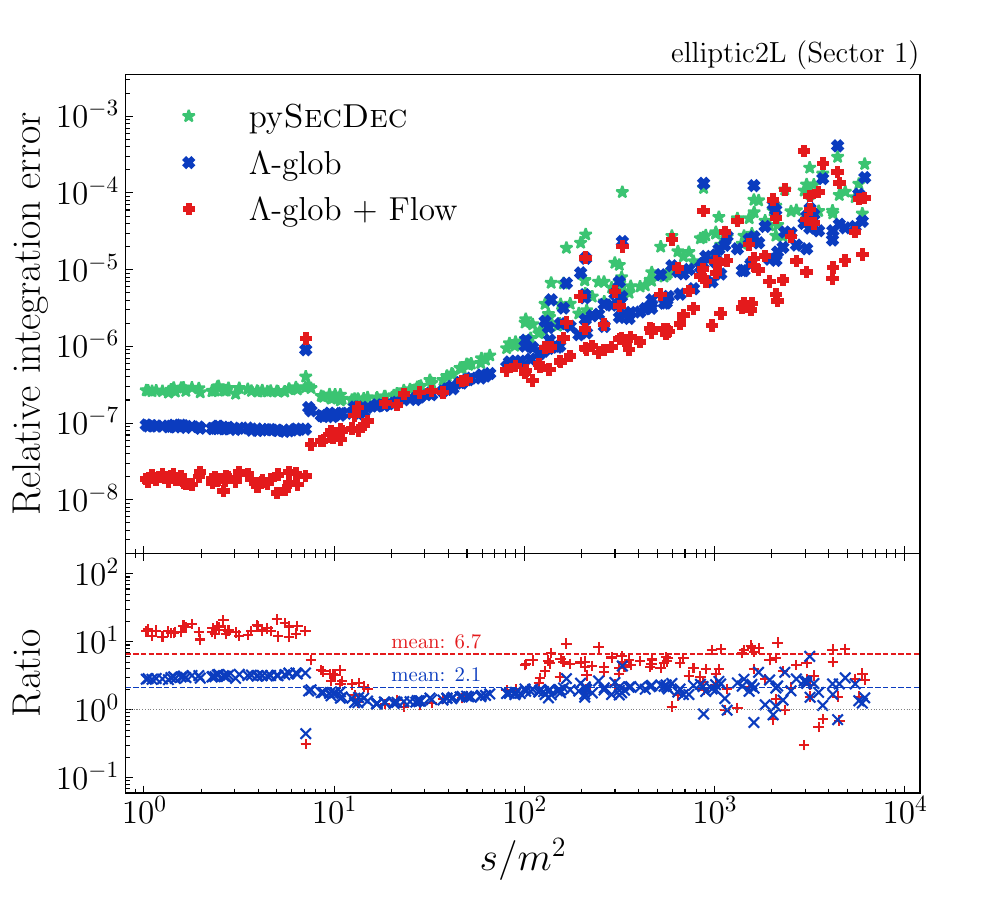}
  \caption{Relative integration error for sector one of the triangle2L (left) and elliptic2L (right) integrals using the standard \pysecdec{} algorithm (green), the $\Lambda$-glob algorithm (blue) and $\Lambda$-glob with additional normalizing flow (red). The lower panel shows the ratios to the standard method.}
  \label{fig:triangle_elliptic}
\end{figure}

In \ref{fig:triangle_elliptic} we show results for the triangle2L (left) and the elliptic2L (right) integral. For both integrals we consider the first sector integral after sector decomposition. We sample 100 phase space points varying over 4-5 orders of magnitude in the squared center-of-mass energy $s\equiv(p_1+p_2)^2$. For both processes, we intentionally consider points below and above threshold, to compare the performance when no contour deformation is needed. We normalized the kinematic invariants using $m^2=1$. For the triangle2L integral, shown in the left panel of \ref{fig:triangle_elliptic}, the average integration error over all phase-space points is reduced by a factor two for the $\Lambda$-glob algorithm and by a factor of 5 for our ML-approach. In the low-energy regime the error reduction stays around the average value. For increasing energies towards threshold at $s/m^2=1$, the absolute integration error of the standard \pysecdec{} method and the pure $\Lambda$-glob algorithm increase, while absolute integration error of our ML-approach keeps decreasing. This results in a relative performance gain by a factor of up to 30 close to the threshold. The threshold being located at  $s/m^2=1$ is a consequence of considering sector one, which effectively corresponds to a topology where one of the massive triangle propagators connecting to $p_3$ is pinched.
In contrast, in the elliptic2L sector 1 integral, shown in the right panel of \ref{fig:triangle_elliptic}, the importance sampling through the normalizing flow reduces the integration error by a factor of 20 and does not show the rising profile towards the threshold. The average integration error is reduced by a factor of 7 or 2 depending on whether the additional mapping of the normalizing flow is used or not. The kinematic points for this diagram are chosen to have varying values of $t=(p_1+p_3)^2$ and $p_4^2$.

In general, for energies close but above threshold the performance gain is less pronounced, as the contour deformation in this regime has less freedom for optimization and the effect of modifying the real parts is diminished.

For increasing energies, the absolute integration error also increases and eventually starts fluctuating. This is driven by the singularities moving toward the endpoints. A possible way to control this behavior has been proposed in Ref.~\cite{Heinrich:2021dbf}. Together with the absolute integration error, the improvement factors also start to fluctuate strongly for large energies.

\begin{figure}[t]
  \includegraphics[page=1,width=0.495\textwidth]{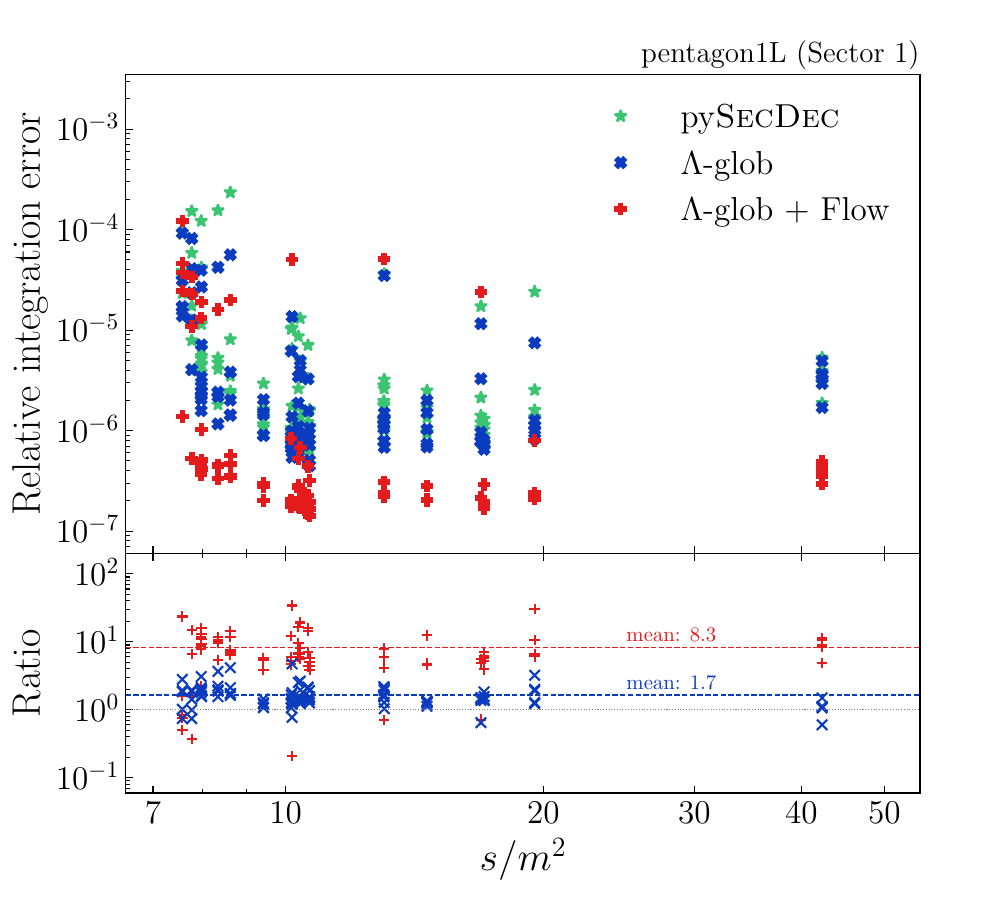}
  \includegraphics[page=1,width=0.495\textwidth]{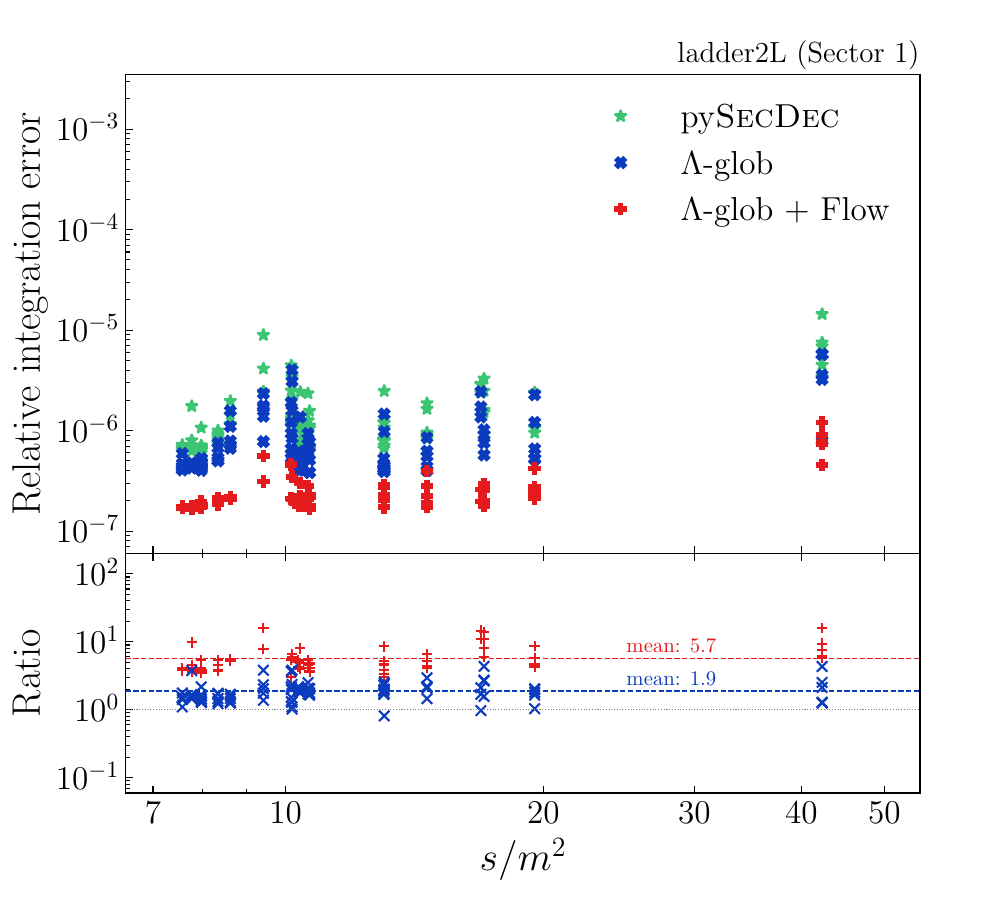}
  \caption{Relative integration error for sector one of the pentagon1L (left) and ladder2L (right) integrals using the standard \pysecdec{} algorithm (green), the $\Lambda$-glob algorithm (blue) and $\Lambda$-glob with additional normalizing flow (red). The lower panel shows the ratios to the standard method.}
  \label{fig:ladder2g_tth}
\end{figure}

Finally, in \ref{fig:ladder2g_tth} we show the results for the more complicated pentagon1L (left) integral and the ladder2L (right) integral. Again, for both integrals we consider the first sector integral after sector decomposition. The increasing complexity originates from both a higher-dimensional integration space, \ie more Feynman parameters, and from having more kinematic scales involved. In order to cover possible dependencies on other kinematic variables than $s$ and $m^2$, we decided to sample different kinematic phase-space for the same values of $s/m^2$. For both integrals we find that the average integration error reduces by a factor two for the $\Lambda$-glob algorithm. By employing the ML-method we achieve an average error reduction factor of $6$ and $8$ for the ladder2L and pentagon1L, respectively. For individual phase-space points we achieve an improvement factor of up to 30. However, there are also phase-space points for which both the $\Lambda$-glob and the flow supplemented algorithm show inferior performance. This clearly indicates the shortcomings of the optimization procedures which are related to the strict sign requirement on the imaginary part.

\section{Outlook}
\label{sec:outlook}

We have shown, for the first time, that the application of modern machine learning methods to numerical multi-loop calculations can lead to a considerable reduction of the numerical uncertainties and hence speed. This has been achieved in a two-step procedure, first applying an algorithm to globally optimize the contour deformation parameters $\lambda$, and subsequently employing a normalizing flow to optimize the complex integration contour, after splitting the full contour deformation into a real and an imaginary part. We have demonstrated the performance with several one- and two-loop examples. All of these examples contain massive propagators and several kinematic scales, leading to a complicated threshold structure of the integrand, such that the contour deformation is a highly non-trivial task, which was dealt with successfully by the neural networks. While the results presented in this paper can only be a first step, they very much motivate further investigations.

\begin{center} \textbf{Acknowledgements} \end{center}

We would like to thank Margarete M\"uhlleitner for useful discussions.
RW acknowledges support by HeiKA and by FRS-FNRS (Belgian National Scientific Research Fund) IISN projects 4.4503.16. The research of AB, GH and TP is supported by the Deutsche Forschungsgemeinschaft (DFG, German Research Foundation) under grant 396021762 — TRR 257 \textsl{Particle Physics Phenomenology after the Higgs Discovery}. This work was supported by the Deutsche Forschungsgemeinschaft (DFG, German Research Foundation) under Germany's Excellence Strategy EXC 2181/1 - 390900948 (the Heidelberg STRUCTURES Excellence Cluster).

\bibliography{literature}
\end{document}